\documentclass[]{aastex631}

\usepackage{graphicx}
\usepackage{amsmath,amssymb,epsfig}
\usepackage{paralist}
\usepackage{comment}

\renewcommand{\vec}[1]{\boldsymbol{\mathrm{#1}}}

\let\Re\relax
\DeclareMathOperator{\Re}{Re}
\begin{document}

\title{Evolving morphology of resolved stellar Einstein rings}

\author[0000-0003-4255-9497]{Slava G. Turyshev}
\affiliation{Jet Propulsion Laboratory, California Institute of Technology,\\
4800 Oak Grove Drive, Pasadena, CA 91109-0899, USA}

\author[0000-0003-3651-9843]{Viktor T. Toth}
\affiliation{Ottawa, Ontario, Canada}

\begin{abstract}

We consider strong gravitational lensing by nearby stars. Using our wave-optical treatment of lensing phenomena, we study Einstein rings that may form around nearby stellar lenses. 
It is remarkable that these rings are bright and large enough to be detected and resolved by existing instruments. Such lensing events have durations of hours or days, with peak light amplification lasting for several minutes. Many such events may be predicted using the Gaia astrometric catalogue. Serendipitous discoveries are also possible. Fortuitous alignments can be used to confirm or discover and study exoplanets. For lenses that have dense stellar regions in their background, these events may occur several time a year, warranting their continuous or recurrent monitoring. Resolved imaging and spectroscopy of the evolving morphology of an Einstein ring offers knowledge about both the lens and the source. The angular size of the Einstein ring amounts to a direct measurement of the lens mass. The changing orientation of the major and minor images of the source offers astrometric information related to the mutual orientation of the objects. The event duration, when the full ring is present, helps to determine the source's size. The sky position of planetary lensing events constrains the planet's orbit. Spectroscopy of the ring allows for direct investigations of the source. The frequency and predictability of these events and the wealth of information that can be obtained by imaging motivate observational campaigns using existing facilities and/or construction of new instruments dedicated to the search and study of Einstein rings that are forming around nearby stars. As a specific example, we consider a predicted 2028 lensing of a red giant by $\alpha$ Centauri A and discuss the relevant science campaign.

\end{abstract}


\section{Introduction}

The gravitational field of a massive body acts as a lens, focusing radiation that passes though it \citep{Einstein:1936}. The sheer size of the resulting gravitational lenses gives rise to impressive optical properties, including extreme light amplification and angular resolution
\citep{Liebes:1964,Paczynski:1986,Paczynski:1996a,Gould:1997}. Depending on its mass distribution, a gravitational lens could form an elaborate caustic structure \citep{Turyshev-Toth:2017,Turyshev-Toth:2021-multipoles,Turyshev-Toth:2021-imaging,Turyshev-Toth:2021-STF-moments}, yielding images of Einstein rings, arcs, crosses and more \citep{Turyshev-Toth:2020-extend,Turyshev-Toth:2021-quartic}  that are both intriguing and informative.  Gravitational lensing is used to probe the distribution of matter in galaxies and clusters of galaxies, to study the distribution of dark matter, to search for exoplanets, etc., e.g.,  \citep{Gaudi:2012,Mao:2012}. Such lenses are used as gravitational telescopes to view the most distant galaxies and the sources of cosmic hydrogen reionization, or study supermassive black holes, even individual stars in serendipitous cases, at cosmological distances.

To be observed, an Einstein ring has to be large enough and bright enough to be visible next to the lens, and also occurring on practically accessible time scales. Einstein rings (or partial arcs, which appear when the lens and source are imperfectly aligned) are most easily seen at extragalactic distances, with entire galaxies or clusters of galaxies acting as lenses \citep{Zwicky:1937,Schneider-Ehlers-Falco:1992}. Such galactic lenses are permanent fixtures in the sky on human timescales.

Smaller objects that are closer to us, such as nearby stars, can also act as gravitational lenses. Lensing events produced by such lenses are transient, shorter in duration due to the large proper motion of the lens across the sky. One manifestation of this phenomenon is microlensing: the intensification of light received from an object aligned behind the lensing mass  \citep{Liebes:1964,Paczynski:1986,Paczynski:1996a,Gould:1997}. These events typically unfold on a timescale of days or weeks, with periods of peak light amplification often measured in mere minutes. However, if the lens passes in front of a dense or otherwise bright stellar background, the formation of either partial or even complete Einstein rings may be observed often.

Consider lensing by nearby objects, namely stars in the Milky Way, especially those in the neighborhood of the solar system. For a lens star at the distance $\overline z$ from us, the angular radius of an Einstein ring formed around it from the light received from a source at large distance $z_0$ from the lens such that $\overline z \ll z_0$
is well known and is given as\footnote{This expression and the relevant results can be easily generalized on the case when $\overline z \sim z_0$ by introducing $1/\overline z \rightarrow 1/\tilde z= 1/\overline z+1/ z_0$ and introduce both $\overline z$ and $z_0$ in all expressions, as was done in \citep{Turyshev-Toth:2019-extend}. For cases considered here $\overline z \ll z_0$, thus the form of expression (\ref{eq:alpha-mu}).  }   {}
\begin{eqnarray}
\theta_{\tt E}&\simeq&\sqrt{\frac{2r_g}{\overline z}}=28.54
\Big(\frac{M_{\tt L}}{M_\odot}\Big)^\frac{1}{2}\Big(\frac{10 ~{\rm pc}}{\overline z}\Big)^\frac{1}{2}~~~{\rm mas},
  \label{eq:alpha-mu}
\end{eqnarray}
where $r_g=2GM_{\tt L}/c^2$ is the Schwarzschild radius of the lens with mass $M_{\tt L}$ and ${\overline z}$ is its distance from us.

Expression (\ref{eq:alpha-mu}) offers good guidance as to where to look for an Einstein ring around any particular star.  As we shall see in this paper, for many stars in our stellar system's neighborhood, the angular radius of the Einstein ring is several tens of milliarcseconds (mas), which is large enough to be resolvable by existing astronomical instruments at optical and even near-IR wavelengths. In those cases, detailed observations of the temporal evolution of the shape and intensity of the Einstein ring, and the orientation of partial arcs that form before or after the main event can provide important clues about the source, the lens and mass distribution in the vicinity of the lens.

Lensing events by nearby stars are typically hard to predict, as they require knowledge of the precise alignments between the lens and source which requires advanced astrometric catalogues. On the other hand, when stars that can serve as potential lenses are located in front of a background that is a dense stellar neighborhood, even unpredicted lensing events may occur with sufficient frequency to make it worthwhile to monitor them for such occurrences.

Among the predicted lensing events, one stands out in particular: in May 2028, our neighbor $\alpha$ Centauri A may briefly align with a distant supergiant \citep{Kervella2016}. If this event happens, existing astronomical facilities can be expected to resolve the Einstein ring even against the bright light of the lens ($\alpha$ Cen A is slightly brighter than our Sun). Given the proximity of the lens and the comparatively large size of the resulting Einstein ring, this event may well serve as a prototype for such observations in the future (like those discussed in \citep{Mustill-etal:2018} in the context of exoplanetary search).

In recent years, we developed a comprehensive wave-theoretical treatment of gravitational lensing by a compact extended lens \citep{Turyshev-Toth:2017,Turyshev-Toth:2019-extend,Turyshev-Toth:2019-blur,Turyshev-Toth:2019-image,Turyshev-Toth:2021-multipoles,Turyshev-Toth:2021-imaging,Turyshev-Toth:2021-quartic,Turyshev-Toth:2021-caustics}
in the weak field limit of the theory of gravitation. We can model and simulate high-resolution lensed images of the Einstein crosses, rings, arcs and more \citep{Turyshev-Toth:2020-extend,Turyshev-Toth:2021-quartic,Turyshev-Toth:2021-STF-moments}. The formalism was primarily intended to characterize the solar gravitational lens, but it is just as applicable to other compact objects.

Using our tools, we can offer quantitative answers to questions such as \begin{inparaenum}[i)]
\item How bright are Einstein rings (or partially formed Einstein rings) that appear around nearby stars? Are they detectable? Does observation require occulting the lens star (using a coronagraph or starshade) or are the Einstein rings sufficiently bright to be visible even against the lens star's glare?
\item How often do such lensing events likely occur? Are these extraordinarily rare events or do they occur with sufficient frequency to make it worthwhile looking for them?
\item What is the expected duration of a typical lensing event?
\item Last but not least, what can we learn by observing such Einstein rings about the background object, the light of which is focused and amplified by the lens?
\end{inparaenum}
In this paper we address the questions above.

The paper is organized as follows: In Section~\ref{sec:aintro} we discuss the optical properties of stellar gravitational lenses and their  imaging capabilities. We assess the power density of the signal received at the focal plane of an imaging telescope, from light emitted by an extended, luminous source viewed through a foreground gravitational lens. In Section~\ref{sec:ER-image} we evaluate the size and brightness of Einstein rings that may be formed around stars in our neighborhood. We consider the evolving morphology of Einstein rings as they form around such stellar lenses. We also estimate the frequency of occurrence of such lensing events using, as an example, a Sun-like star in the foreground, with a dense star cluster in the distant background. In Section~\ref{sec:acent} we apply our understanding to the case of the anticipated lensing event by $\alpha$ Cen A in 2028 and also model a hypothetical planet. Finally, in Section~\ref{sec:disc} we  present and discuss our conclusions.

\section{Fundamentals}
\label{sec:aintro}

We consider strong gravitational lensing. In addition to the photometric variability in the brightness of a source (which can be explored through microlensing observations), we quantify the structure and evolving morphology of the images formed by gravitational lens, characterizing the event using a wave-theoretical formalism.

\subsection{Lensing geometry}
\label{sec:geom}

In Refs.~\citep{Turyshev-Toth:2019-extend,Turyshev-Toth:2020-extend,Turyshev-Toth:2019-blur,Turyshev-Toth:2019-image,Turyshev-Toth:2021-multipoles,Turyshev-Toth:2021-imaging,Turyshev-Toth:2021-quartic,Turyshev-Toth:2021-caustics}, while dealing with the solar gravitational lens, we developed a comprehensive wave-optical formalism to describe gravitational lensing by a realistic extended mass distribution, characterized by an infinite series of mass multipole moments. This approach allows us to explore the theory beyond microlensing \citep{Liebes:1964,Paczynski:1986,Schneider-Ehlers-Falco:1992}, by providing a wave-optical treatment of the gravitational lensing phenomena by any extended compact lens.  The theory describes prospective observations that include photometry, imaging and spectroscopy.

We describe a lensing configuration that consists of an extended source, a compact lens, and an observer. The three are approximately aligned along a straight line at distances $z_0$ and $\overline z$ from source to lens and lens to observer, respectively\footnote{In the microlensing literature, these quantities are commonly denoted either as  ${\overline z}=D_{\rm d}$ and $z_0=D_{\rm ds}=D_{\rm s}-D_{\rm d}$  (see \citep{Schneider-Ehlers-Falco:1992}) or ${\overline z}=D_l$ and $z_0=D_{ls}=D_s-D_l$ (as, for instance, in \citep{Gould:1996,Gaudi:2012}). In the present paper, we follow the formalism that we developed for the solar gravitational lens, while working in a lens-centric coordinate system, e.g., as discussed in \citep{Turyshev-Toth:2019-extend}.}. The scenarios of resolved imaging that we presently consider assume that the lens is quite close to the observer, with typical imaging configurations characterized as $\overline z\ll z_0$.

\begin{figure}[t!]
\begin{center}
\includegraphics[scale=0.7]{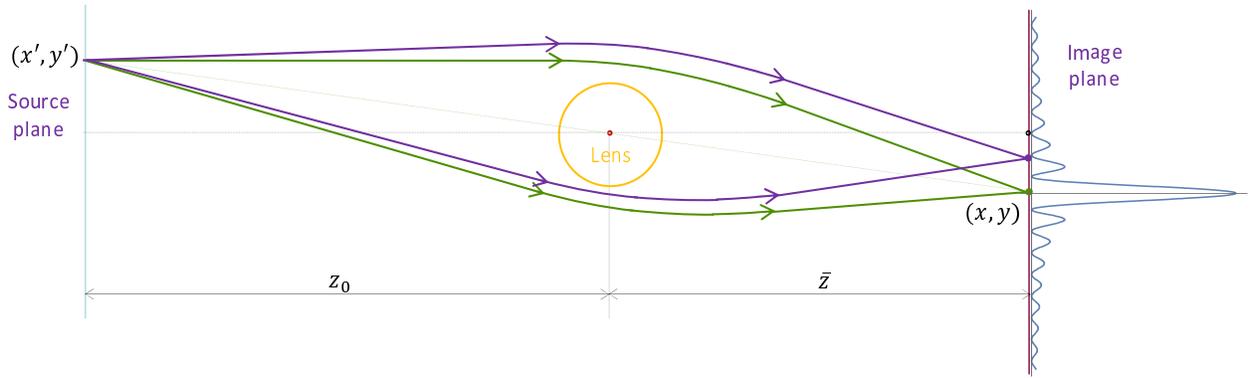}
\end{center}
\caption{\label{fig:imaging-geom}The geometry of imaging a point source with a stellar gravitational lens. A point source with coordinates $(x',y')$ is positioned in the source plane, at the distance $z_0$ from the lens. The  image plane is at the distance ${\overline z}$ from the lens. Rays with different optical paths produce a diffraction pattern $\propto J^2_0\big(k\sqrt{{2r_g}/{\overline z}} |{\vec x}+({\overline z}/{{ z}_0}){\vec x'}|\big)$ in the image plane, see \citep{Turyshev-Toth:2020-extend}.}
\end{figure}

We consider an extended source of light with radius $R_{\tt s}$ at a finite distance, $z_0$, from the lens. We use a lens-centric cylindrical coordinate system $(\rho,\phi,z)$, with the $z$-axis corresponding to the primary optical axis: a line connecting the center of the source to the that of the lens \citep{Turyshev-Toth:2020-extend}, as shown in Fig.~\ref{fig:imaging-geom}; thus, the primary optical axis  is defined by its directional vector $\vec k=(0,0,1)$ in the chosen coordinate system.
Furthermore, to discuss imaging, we introduce the source plane, image plane, and the optical telescope's focal (or sensor) plane, each orthogonal to ${\vec k}$.

Next, we introduce two-dimensional coordinates to describe points in the source plane, $\vec x'$; the position of the telescope in the image plane,  $\vec x_0$; points in the image plane within the telescope's aperture, $\vec x$; and points in the optical telescope's focal plane ${\vec x}_i$:
\begin{eqnarray}
\{{\vec x}'\}&\equiv& \rho'\big(\cos\phi',\sin\phi'\big)=\rho'{\vec n}',
\label{eq:x'}
\qquad \{{\vec x}_0\}~\equiv~ \rho_0\big(\cos\phi_0,\sin\phi_0\big)=\rho_0{\vec n}_0,
\label{eq:x0}\\
\{{\vec x}\}&\equiv&
\rho\big(\cos\phi,\sin\phi\big)=\rho\,{\vec n},
\label{eq:x}
\qquad\quad~
 \{{\vec x}_i\} ~\equiv~
 \rho_i\big(\cos\phi_i,\sin\phi_i\big)=\rho_i{\vec n}_i.
  \label{eq:p}
\end{eqnarray}

\begin{figure}
\begin{center}
\includegraphics[scale=0.9]{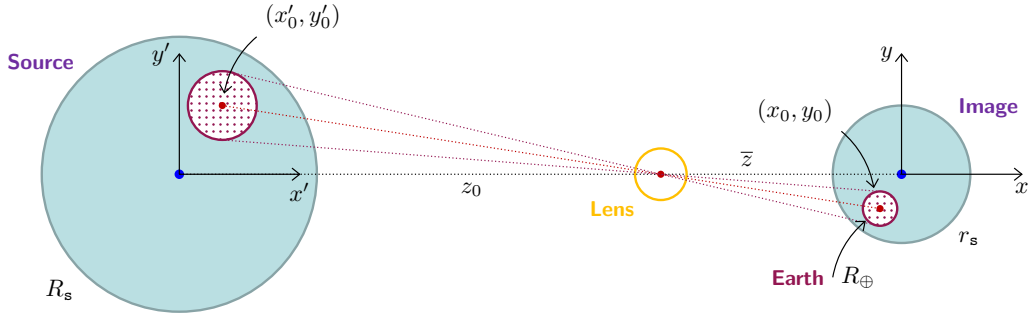}
\end{center}
\caption{\label{fig:dir-image}Imaging of a large, extended source by a stellar gravitational lens. The Earth is situated in the image plane. In the depicted example, the projected image of the source is larger than the Earth, $r_{\tt s}=({\overline z}/z_0)R_{\tt s}> R_\oplus$. A network of ground based telescopes on the Earth and those in space travels through this image during a lensing event. A smaller or more distant source would project an image smaller than the Earth, which would travel across the Earth's surface during the lensing event.}
\end{figure}

A stellar gravitational lens projects light by focusing in the opposite quadrant in the image plane (see Fig.~\ref{fig:dir-image}) while also reducing the size of the resolved image compared to the source by a factor of
{}
\begin{eqnarray}
\beta=\frac{\overline z}{z_0}\sim1.25\times 10^{-3}\,\Big(\frac{\overline z}{10 ~{\rm pc}}\Big) \Big(\frac{8~{\rm kpc}}{z_0}\Big).
  \label{eq:beta}
\end{eqnarray}
For a source with radius $R_{\tt s}$, positioned at a distance of $z_0$ from the lens, the projected image of this target within the solar system that is at the distance ${\overline z}$ from the lens will have the diameter \citep{Turyshev-Toth:2020-extend}
{}
\begin{equation}
d_{\tt s}=2r_{\tt s}=\frac{\overline z}{z_0}2R_{\tt s}=1739.25\,\Big(\frac{R_{\tt s}}{R_\odot}\Big)\Big(\frac{\overline z}{10 ~{\rm pc}}\Big) \Big(\frac{8~{\rm kpc}}{z_0}\Big)~{\rm km}.
\label{eq:rE}
\end{equation}

The position of this image is not static. It sweeps through the solar system as a result of the combined motion of the solar system, the lens, and (to a lesser extent if ${\overline z}\ll z_0$) the source. A telescope that is inside the area defined by the projected image, looking back at the lens, can view the the Einstein ring that forms around the lens from light received from the source. Meanwhile, photometric measurements of the light received by the telescope amount to sampling the projected image at the telescope's location. As the telescope moves across the image (or equivalently, the image moves with respect to the telescope's location), the telescope's path across the image provides a one-dimensional cross section that can reveal details about the source. Multiple telescopes at different geographic locations on the Earth, perhaps augmented by space-based instruments, may trace different, approximately parallel paths, extending this scanning to two dimensions, providing a means to reconstruct a resolved image of the source or at least a part of the source (see Fig.~\ref{fig:dir-image}).

\subsection{Image formation}
\label{sec:im-form-proc}

\begin{figure}
\begin{center}
\includegraphics[scale=0.65]{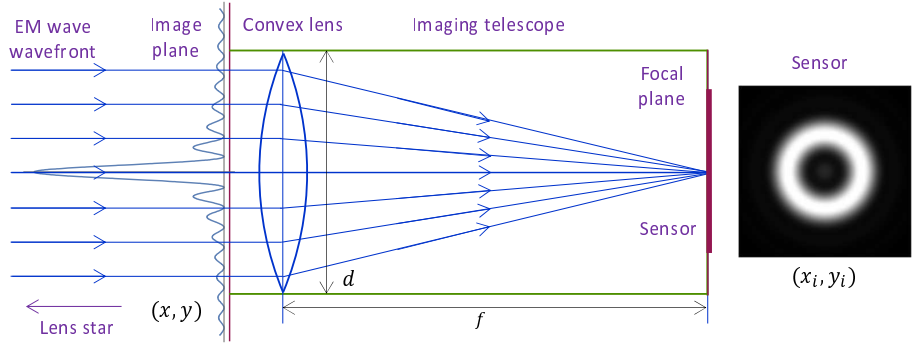}
\end{center}
\caption{\label{fig:imaging-sensor} Imaging a point source with a stellar gravitational lens using a telescope. The telescope is positioned on the optical axis that connects the source and the lens and it ``sees'' the full Einstein ring. The telescope is represented by a convex lens with a diameter $d$ and a focal length $f$. Positions in the  image plane, $(x,y)$, and on the image sensor, $(x_i,y_i)$, that is positioned at the optical telescope's focal plane are also shown.}
\end{figure}

To model the imaging of distant objects with a stellar gravitational lens, we represent an imaging telescope by a convex lens with aperture $d$ and focal distance $f$; see Fig.~\ref{fig:imaging-sensor}. We position the telescope at a point with coordinates ${\vec x}_0$ in the image plane.
We rely on definitions (\ref{eq:x'})--(\ref{eq:p}), but slightly redefine them  by introducing ${\vec x}_0=-\beta{\vec x}_0'$, which allows us to define coordinates ${\vec x}''$ on the source plane that are centered on ${\vec x}_0'$:
{}
\begin{eqnarray}
{\vec x}''={\vec x}'-{\vec x}'_0\equiv \rho''{\vec n}''= \rho''(\cos\phi'',\sin\phi'').
  \label{eq:coord2}
\end{eqnarray}

With these definitions, following the procedures outlined in \citep{Born-Wolf:1999,Turyshev-Toth:2019-extend,Turyshev-Toth:2020-extend}, the intensity distribution on the image sensor of the imaging telescope corresponding to a signal received from a distant source is given as
  {}
\begin{eqnarray}
I_{\tt }({\vec x}_i,{\vec x}_0) &=& \frac{\mu_0}{z_0^2} \Big(\frac{kd^2}{8f}\Big)^2\hskip-4pt
\iint d^2{\vec x}''  B_{\tt s}(\vec x'', \lambda)  {\cal A}^2({\vec x}_i,{\vec x}''),
  \label{eq:pow-blur}
\end{eqnarray}
where $k=2\pi/\lambda$ is the wavenumber of an EM wave, $\mu_0=2\pi kr_g\simeq1.17\times 10^{11}\,
({M_{\tt L}}/{M_\odot})({1\,\mu{\rm m}}/{\lambda})$ is the light amplification factor of a stellar gravitational lens,  $B_{\tt s}(\vec x'', \lambda) $ is the source's surface brightness,  and ${\cal A}^2({\vec x}_i,{\vec x}'')$ is the Fourier-transformed squared amplitude derived from the point-spread function (PSF) of the lens. The functional form of the PSF in these regions and the Fourier-transformed amplitude ${\cal A}({\vec x}_i,{\vec x}'')$ are well established (see \citep{Turyshev-Toth:2019-extend,Turyshev-Toth:2020-extend,Turyshev-Toth:2021-all-regions} for details).

\begin{figure}
\begin{center}
\includegraphics[scale=0.85]{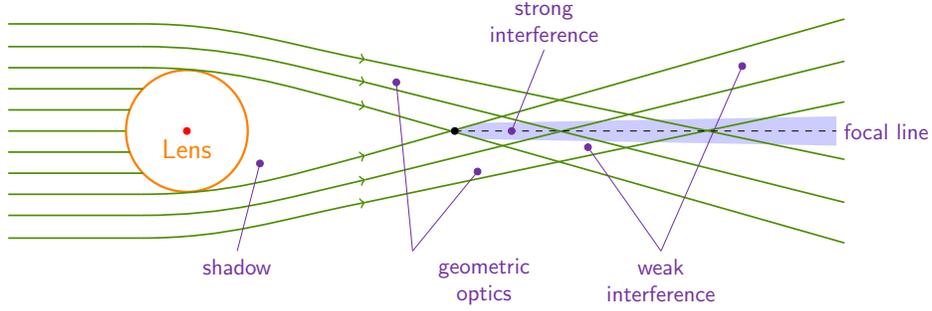}
\end{center}
\caption{\label{fig:regions}The different imaging regions that form behind a gravitational lens. Adapted from \citep{Turyshev-Toth:2019-image}. In the geometric optics region, a single (possibly deflected, amplified) image of the source is visible. In the weak interference region, two images are seen that may be inequal in brightness and distance from the lens. In the strong interference region the two images, now equal in brightness, widen into arcs and ultimately form the Einstein ring when viewed from the immediate vicinity of the optical axis.}
\end{figure}

The quantity (\ref{eq:pow-blur}) represents the intensity of light deposited on the image sensor; in other words, this is the image that we seek to detect and study. Depending on the separation from the primary optical axis, $\rho_0$, there are two lensing regions formed by a lens, including the strong and weak interference regions; for comparison purposes, we also consider the geometric optics region characterized by gravitational bending of light, and the free-space region where no lensing is present (see Fig.~\ref{fig:regions}). Based on our prior work  \citep{Turyshev-Toth:2017,Turyshev-Toth:2019-extend,Turyshev-Toth:2020-extend,Turyshev-Toth:2019-image,Turyshev-Toth:2021-all-regions}, for a point source we characterize these regions as follows:
\begin{itemize}
\item  At a very large distance from the primary optical axis the presence of the lens can be neglected. A diffraction-limited telescope illuminated by a compact, unresolved source forms an image in its image plane in the form of the Airy disk corresponding to an unamplified image of the source. The Fourier-transformed amplitude, $ {\cal A}_{\tt 0}({\vec x}_i, {\vec x}_0)$, for this region is well-established in the literature (e.g., \citep{Abramovitz-Stegun:1965,Turyshev-Toth:2021-imaging}).

\item Approaching the optical axis, light from the distant source is deflected by the lens. We enter the region of geometric optics.

\item Further approaching the optical axis, we enter the region where light from the source becomes amplified by the lens. As we reach the region where $\sqrt{\lambda \overline z/\pi} \lesssim \rho_0\lesssim \sqrt{2r_g \overline z} - R_{\tt L}$, with $R_{\tt L}$ being the radius of the lens, a minor image appears from behind the lens on the opposite side from the major image. This characterizes the region of weak interference. In Refs. \citep{Turyshev-Toth:2019-extend,Turyshev-Toth:2020-extend,Turyshev-Toth:2021-imaging,Turyshev-Toth:2021-all-regions}), we developed a wave-optical description of image formation in this region. The Fourier-transformed amplitude, $ {\cal A}_{\tt weak.int.}({\vec x}_i, {\vec x}_0)$,  for this region is also well known, e.g., \citep{Turyshev-Toth:2020-extend,Turyshev-Toth:2021-imaging,Turyshev-Toth:2021-all-regions}. This formalism also describes the geometric optics region, with the minor image removed to account for the opaque lens.

\item As $\rho_0$ continues to decrease, the major and minor images become equal in magnitude. They settle on the circumference of the yet-to-form Einstein ring. The two images continue to increase in brightness but remain unresolved so long as $\lambda/\theta_{\tt E} \ll \rho_0 \lesssim \sqrt{\lambda \overline z/\pi} $. Eventually, as $\rho_0\to 0$, the major and minor images widen into arcs and finally form the Einstein ring. This is the ``pencil sharp'' region of strong interference \citep{Turyshev-Toth:2017}. The Fourier-transformed amplitude, $ {\cal A}_{\tt ER}({\vec x}_i, {\vec x}_0)$,  for this region was derived in  \citep{Turyshev-Toth:2020-extend,Turyshev-Toth:2021-imaging,Turyshev-Toth:2021-all-regions}.
\end{itemize}

The approximation that we developed for the weak interference region becomes valid when $\rho_0\gg \lambda/\theta_{\tt E}$ and, except for the need to remove the minor image in the geometric optics region, remains valid at arbitrarily large distances from the optical axis. This approximation properly predicts the displacement and light amplification of the major and minor images, but not the formation of the arcs that merge into Einstein ring. In contrast, the strong interference region approximation remains valid for so long as the displacement of the major and minor images from the Einstein ring can be ignored. In most practical imaging scenarios, there is substantial overlap, spanning several orders of magnitude between these regions of validity. This can be advantageous, as the weak interference region approximation is significantly less costly to compute.

An extended source can be modeled as a collection of point sources, integrating over the surface area of the source. In the special case when the source can be modeled as a uniformly illuminated disk, this integration is greatly simplified, leading to useful expressions for the weak and strong interference regions, which we now present in more detail.

\subsubsection{Image formation in free space, the geometric and the weak interference regions}
\label{sec:weak-int}

In the case when the influence of the lens may be neglected, the solution for the EM field is derived using free-space EM wave propagation. A solution for the EM field for the weak interference region consists of a combination of the gravity-modified incident wave and also the scattered wave that results from the diffraction of the incident wave on the solar gravity field \citep{Turyshev-Toth:2017}. Following the approach presented in  \citep{Turyshev-Toth:2019-extend,Turyshev-Toth:2020-extend,Turyshev-Toth:2019-image}, we use the method of stationary phase to develop solutions for Maxwell's equations in the weak field approximation to describe the incident and scattered EM fields and derive the expression to determine the intensity distribution for the signals received in these two regions.

For large physical separations from the optical axis, the quantity ${\cal A}^2({\vec x}_i,{\vec x}_0)$, needed in (\ref{eq:pow-blur}), for both regions---the region with no gravitational lensing, ${\cal A}^2_{\tt 0}({\vec x}_i, {\vec x}_0)$, and the weak interference region, ${\cal A}^2_{\tt weak.int.}({\vec x}_i,{\vec x}_0)$---is given as
 {}
\begin{eqnarray}
  {\cal A}^2_{\tt 0}({\vec x}_i, {\vec x}_0)&=&
\bigg(\frac{
2J_1\big(kd\, \hat v({\vec x}_i,{\vec x}_0)\big)}{kd\, \hat v({\vec x}_i,{\vec x}_0)}\bigg)^2,
  \label{eq:FI-go-mu}\\
{\cal A}^2_{\tt weak.int.}({\vec x}_i,{\vec x}_0)&=&
a^2_{\tt in}\bigg(\frac{
2J_1\big(kd\, \hat v_+({\vec x}_i,{\vec x}_0)\big)}{kd\, \hat v_+({\vec x}_i,{\vec x}_0)}\bigg)^2+a^2_{\tt sc}\bigg(\frac{
2J_1\big(kd\, \hat v_-({\vec x}_i,{\vec x}_0)\big)}{kd\, \hat v_-({\vec x}_i,{\vec x}_0)}\bigg)^2+{\cal O}(r^2_g),~~~
  \label{eq:FI-ir+-mu}
\end{eqnarray}
where $J_1(x)$ is the Bessel function of the first kind. The real-valued factors $a^2_{\tt in/sc}$ are the amplification factors for the incident and scattered waves  (corresponding to the major and minor images known from the geometric-optics theory of microlensing \citep{Liebes:1964,Paczynski:1986,Schneider-Ehlers-Falco:1992}) that are given as (with the upper sign is for $a^2_{\tt in}$ and the lower index is for $a^2_{\tt sc}$)
{}
 \begin{eqnarray}
a^2_{\tt in/sc} &=&
\frac{\big[{\textstyle\frac{1}{2}} (\sqrt{1+{8r_g {\overline z}}/{\rho_0^2}}\pm1)\big]^2}{\sqrt{1+{8r_g {\overline z}}/{\rho_0^2}}}\equiv \frac{[{\textstyle\frac{1}{2}} \big(\sqrt{u^2+4}\pm  u\big)]^2}{u\sqrt{u^2+4}},\qquad {\rm where}\qquad u=\frac{\theta}{\theta_{\tt E}}, \qquad
\theta=\frac{\rho_0}{\overline z}.
    \label{eq:a_insc}
\end{eqnarray}

It is convenient to introduce $ \theta_\pm$ for the angles corresponding to the positions of the observed major and minor images \citep{Liebes:1964,Schneider-Ehlers-Falco:1992}: $ \theta_\pm= {\textstyle\frac{1}{2}} \big(\sqrt{\theta^2+4\theta_{\tt E}^2}\pm  \theta\big)\equiv {\textstyle\frac{1}{2}} \big(\sqrt{u^2+4}\pm  u\big)\theta_{\tt E}$, naturally corresponding to similar quantities introduced in microlensing theory \citep{Liebes:1964,Paczynski:1986,Schneider-Ehlers-Falco:1992}.  Next, the normalized spatial frequencies
$\hat v_+({\vec x}_i,{\vec x}_0)$ and $\hat v_-({\vec x}_i,{\vec x}_0)$ are given as
{}
\begin{eqnarray}
    \hat v_+({\vec x}_i,{\vec x}_0)&=&
\Big\{{\textstyle\frac{1}{4}}\Big(\theta_+-\frac{\rho_i}{f}\Big)^2+\theta_+\frac{\rho_i}{f} \cos^2[{\textstyle\frac{1}{2}}\big(\phi_0-\phi_i\big)]\Big\}^\frac{1}{2},
  \label{eq:mono-v+}\\
\hat v_-({\vec x}_i,{\vec x}_0)&=&
\Big\{{\textstyle\frac{1}{4}}\Big(\theta_--\frac{\rho_i}{f}\Big)^2+\theta_-\frac{\rho_i}{f}\sin^2[{\textstyle\frac{1}{2}}\big(\phi_0-\phi_i\big)] \Big\}^\frac{1}{2}.
  \label{eq:mono-v-}
\end{eqnarray}
Also, $  \hat v({\vec x}_i,{\vec x}_0)$ in (\ref{eq:FI-go-mu}) for the unamplified signal is obtained from (\ref{eq:mono-v+}) by setting $r_g\rightarrow 0$, which, in this case,  yields $ \theta_+\rightarrow \theta$, while the amplification factors for the incident and scattered waves from (\ref{eq:a_insc}) behave as $a^2_{\tt in}\rightarrow1$ and $a^2_{\tt sc}\rightarrow0$.

For an actual astrophysical source, $B_s({\vec x}'')$ is, of course, an arbitrary function of the coordinates ${\vec x}'$ and thus the integral (\ref{eq:pow-blur}) can only be evaluated numerically. However, we can obtain an analytic result in the simple case of a disk of uniform brightness, characterized by $B_{\tt s}(\vec x'', \lambda)=B_{\tt s}(\lambda)$. As a result, from (\ref{eq:pow-blur}) with (\ref{eq:FI-go-mu}), (\ref{eq:FI-ir+-mu}) we determine the following intensities to be observed in the focal plane of the imaging telescope:
 {}
\begin{eqnarray}
I_{\tt 0}({\vec x}_i,{\vec x}_0,\lambda)&=&\frac{B_{\tt s}(\lambda)}{z_0^2}
 \Big(\frac{kd^2}{8f}\Big)^2 \pi R^2_{\tt s}
 \bigg(\frac{
2J_1\big(kd\, \hat v({\vec x}_i,{\vec x}_0)\big)}{kd\, \hat v({\vec x}_i,{\vec x}_0)}\bigg)^2,
  \label{eq:FI-go-Int}\\
I_{\tt weak.int}({\vec x}_i,{\vec x}_0,\lambda)&=&\frac{B_{\tt s}(\lambda)}{z_0^2}  \Big(\frac{kd^2}{8f}\Big)^2 \pi R^2_{\tt s} \bigg\{
a^2_{\tt in} \bigg(\frac{
2J_1\big(kd\, \hat v_+({\vec x}_i,{\vec x}_0)\big)}{kd\, \hat v_+({\vec x}_i,{\vec x}_0)}\bigg)^2+
a^2_{\tt sc} \bigg(\frac{
2J_1\big(kd\, \hat v_-({\vec x}_i,{\vec x}_0)\big)}{kd\, \hat v_-({\vec x}_i,{\vec x}_0)}\bigg)^2+{\cal O}(r^2_g)\bigg\}.~~~~
  \label{eq:FI-ir+-Int}
\end{eqnarray}

Eqs.~(\ref{eq:FI-go-Int})--(\ref{eq:FI-ir+-Int}) represent the intensity distribution of images that is formed on the image sensor of an imaging telescope and describe two different lensing configurations. Exp.~(\ref{eq:FI-go-Int}) describes the free-space light propagation from the source to an observer in the case when one may neglect the presence of the lens. In this case only one spot of light in the form of an Airy pattern is present on the image sensor corresponding to an unamplified image of the source.

The result (\ref{eq:FI-ir+-Int}) describes imaging at a smaller distance from the primary optical axis. It describes two spots of light in the form of the Airy patterns with different amplitudes  corresponding to the incident and scattered waves containing the terms with $\hat v_+$ and $\hat v_-$, correspondingly.  Examining (\ref{eq:FI-ir+-Int}) in conjunction with (\ref{eq:mono-v+})--(\ref{eq:mono-v-}), we see that these expressions nearly vanish for most values of $v_\pm$, except when  $\hat v_\pm$  becomes zero which happens, when $\rho_i/f  \rightarrow \theta_\pm$. When this happens, we observe a spot that is outside the Einstein ring (for $\rho_i/f  \rightarrow \theta_+$) describing the major image and the other one inside the ring (for $\rho_i/f  \rightarrow \theta_-$) describing the minor image. This approach provides a wave-optical treatment for microlensing phenomena that are usually described using the language of geometric optics \citep{Schneider-Ehlers-Falco:1992}.

\subsubsection{Image formation in the strong interference region}
\label{sec:intense-in-m}

To describe the view by an imaging telescope with sufficient resolution, it is necessary to go beyond the microlensing theory \citep{Liebes:1964,Paczynski:1986,Schneider-Ehlers-Falco:1992,Gaudi:2012}, which describes lensing using the geometric optics approximation. In this approach, the total amplification derived from (\ref{eq:a_insc}) as a sum of $(a^2_{\tt in} +a^2_{\tt sc})$ diverges at the principal optical axis, $u=0$. This is where a complete wave-optical description of the strong gravitational lensing is needed. It is here, in the strong interference region light from a compact source forms the image of an Einstein ring \citep{Turyshev-Toth:2017}. Here, we present a brief summary of our recently developed wave-theoretical description (see, e.g., \citep{Turyshev-Toth:2020-extend,Turyshev-Toth:2019-blur,Turyshev-Toth:2019-image}) of the formation of Einstein rings.

In the strong interference region, the telescope's displacement from the primary optical axis is small, $\rho_0 \lesssim \sqrt{\lambda \overline z/\pi} $, comparable in fact in magnitude to $\lambda/\theta_{\tt E}$. The quantity ${\cal A}^2_{\tt ER}({\vec x}_i,{\vec x}'')$  needed to evaluate (\ref{eq:pow-blur}), is given as  (see details in \citep{Turyshev-Toth:2020-extend}):
{}
\begin{eqnarray}
{\cal A}^2_{\tt ER}({\vec x}_i,{\vec x}'')&=&
 \frac{1}{2\pi \alpha\beta\rho''}\Big\{\Big(\frac{2J_1(k d \, \hat u_0(\phi'',\vec x_i))}{k d \, \hat u_0(\phi'',\vec x_i)}\Big)^2+
 \Big(\frac{2J_1(k d \, \hat u_\pi(\phi'',\vec x_i))}{k d \, \hat u_\pi(\phi'',\vec x_i)}\Big)^2+\nonumber\\
 &&\hskip 50pt +~ 2\sin(2\alpha\beta\rho'')\Big(\frac{2J_1(k d \, \hat u_0(\phi'',\vec x_i))}{k d \, \hat u_0(\phi'',\vec x_i)}\Big)\Big(\frac{2J_1(k d \, \hat u_\pi(\phi'',\vec x_i))}{k d \, \hat u_\pi(\phi'',\vec x_i)}\Big)\Big\},
  \label{eq:amp-mono-int2}
\end{eqnarray}
where $\alpha$ is the spatial frequency introduced by  the Einstein ring is given as
{}
\begin{eqnarray}
\alpha=k\theta_{\tt E}\equiv k\sqrt{\frac{2r_g}{\overline z}}=0.869\Big(\frac{1~\mu{\rm m}}{\lambda}\Big)
\Big(\frac{M_{\tt L}}{M_\odot}\Big)^\frac{1}{2}\Big(\frac{10 ~{\rm pc}}{\overline z}\Big)^\frac{1}{2}~~~{\rm m}^{-1},
  \label{eq:alpha}
\end{eqnarray}
the ratio $\beta$ is given by (\ref{eq:beta}) as $\beta=\overline z/z_0$, and
 $\hat u_0(\phi'',\vec x_i)$ and $\hat u_\pi(\phi'',\vec x_i)$ are the normalized spatial frequencies
{}
\begin{eqnarray}
\hat u_0(\phi'',\vec x_i)&=&
\Big\{{\textstyle\frac{1}{4}}\Big(\theta_{\tt E}-\frac{\rho_i}{f}
\Big)^2+\theta_{\tt E}\frac{\rho_i}{f}
\cos^2[{\textstyle\frac{1}{2}}\big(\phi''-\phi_i\big)]
\Big\}^\frac{1}{2},
  \label{eq:mono-u-hat-0+} \\
  \hat u_\pi(\phi'',\vec x_i)&=&
\Big\{{\textstyle\frac{1}{4}}\Big(\theta_{\tt E}-\frac{\rho_i}{f}
\Big)^2+ \theta_{\tt E}\frac{\rho_i}{f}
\sin^2[{\textstyle\frac{1}{2}}\big(\phi''-\phi_i\big)]\Big\}^\frac{1}{2}.
  \label{eq:mono-u-hat-pi+}
\end{eqnarray}

To compute the power received from the resolved source, we introduce a new coordinate system in the source plane, ${\vec x}''$,  given by (\ref{eq:coord2}). As vector ${\vec x}'_0$ is constant, $dx'dy'=dx''dy''$. Next, in the new coordinate system, we use polar coordinates $(r'',\phi'')$. In these coordinates, the circular edge of the source, $R_{\tt s}$, is no longer a circle but a curve, $\rho_{\tt s}(\phi'')$, the radial distance of which is given by the following relation: $\rho_{\tt s}(\phi'') =(R_{\tt s}^2-{\rho'_0}^2\sin^2\phi'')^\frac{1}{2}-\rho'_0\cos\phi''.$

Again, assuming a spatially uniform surface brightness, ${B}_{\tt s}({\vec x}'',\lambda) ={B}_{\tt s}(\lambda)$, for a telescope position within the image we use (\ref{eq:amp-mono-int2}) and integrate (\ref{eq:pow-blur}) as follows:
{}
\begin{eqnarray}
I_{\tt on}({\vec x}_i,{\vec x}_0,\lambda)  &=&\mu_0 \frac{B_{\tt s}(\lambda)}{z_0^2} \Big(\frac{kd^2}{8f}\Big)^2
\int_0^{2\pi} \hskip -6pt d\phi'' \int_0^{\rho_{\tt s}}\hskip -4pt \rho'' d\rho''  {\cal A}^2({\vec x}_i,{\vec x}'') \equiv
\frac{B_{\tt s}(\lambda)}{z_0^2} \Big(\frac{kd^2}{8f}\Big)^2    \frac{ \mu_0R_{\tt s}}{\alpha\beta} \times\nonumber\\
&&\hskip -55pt \times\,
\frac{ 1}{2\pi} \int_0^{2\pi} \hskip -8pt d\phi'' \bigg\{
\Big(\sqrt{1-\Big(\frac{\rho_0}{r_{\tt s}}\Big)^2\sin^2\phi''}-\frac{\rho_0}{r_{\tt s}}\cos\phi''\Big)
\bigg(\Big(\frac{2J_1(k d\, \hat u_0(\phi'',\vec x_i))}{k d\, \hat u_0(\phi'',\vec x_i)}\Big)^2+
 \Big(\frac{2J_1(k d\, \hat u_\pi(\phi'',\vec x_i))}{k d\, \hat u_\pi(\phi'',\vec x_i)}\Big)^2\bigg)+\nonumber\\
 &&\hskip -40pt +~
\frac{ 2}{k r_{\tt s}}\sin^2\Big[\alpha r_{\tt s}
\Big(\sqrt{1-\Big(\frac{\rho_0}{r_{\tt s}}\Big)^2\sin^2\phi''}-\frac{\rho_0}{r_{\tt s}}\cos\phi''\Big)\Big]
\Big(\frac{2J_1(k d\, \hat u_0(\phi'',\vec x_i))}{k d\, \hat u_0(\phi'',\vec x_i)}\Big)\Big(\frac{2J_1(k d\, \hat u_\pi(\phi'',\vec x_i))}{k d\, \hat u_\pi(\phi'',\vec x_i)}\Big)\bigg\}.
  \label{eq:intense2*}
\end{eqnarray}

As discussed in \citep{Turyshev-Toth:2019-blur}, blur from an extended source is present even outside the direct image of the source. Therefore, a telescope positioned at $\rho_0 > r_{\tt s}$ will still receive light from the source. In this case, the blur for the off-image position, $I_{\tt off}({\vec x}_0)$, is obtained by integrating (\ref{eq:pow-blur}) over the surface of the source as it is seen from an off-image coordinate system. In this case, the telescope is  outside the image, thus, the integration limits change. First, we note that the circular edge of the source, $R_{\tt s}$, is given by a curve, $\rho_{\tt s}(\phi'')$, the radial distance of which in this polar coordinate system is given as $\rho^\pm_{\tt s}(\phi'') =\pm (R_{\tt s}^2-{\rho'_0}^2\sin^2\phi'')^\frac{1}{2}+\rho'_0\cos\phi''$, with the angle $\phi''$ in this case is defined so that $\phi''=0$ when pointing at the center of the source. The angle $\phi''$ varies only within the range $\phi''\in [\phi_-,\phi_+]$, with $\phi_\pm=\pm \arcsin ({R_{\tt s}}/{\rho'_0})$. Also, for any angle $\phi''$ there will be two solutions for $\rho_{\tt s}(\phi'')$, given as $(\rho^-_{\tt s},\rho^+_{\tt s})$. Note, that ${\rho'_0}/{R_{\tt s}}\equiv {\rho_0}/{r_{\tt s}}$.

For a spatially uniform surface brightness of the source, $B_{\tt s}(\vec x'',\lambda)=B_{\tt s}(\lambda)$, for an off-image telescope position, we use (\ref{eq:amp-mono-int2}) to integrate (\ref{eq:pow-blur}) as
{}
\begin{eqnarray}
I_{\tt off}({\vec x}_i,{\vec x}_0,\lambda)  &=& \mu_0\frac{B_{\tt s}(\lambda)}{z_0^2} \Big(\frac{kd^2}{8f}\Big)^2 \int_{\phi_-}^{\phi_+} \hskip -8pt d\phi''
\int_{\rho^-_{\tt s}}^{\rho^+_{\tt s}}\hskip -3pt \rho'' d\rho''
  {\cal A}^2_{\tt }({\vec x}_i,{\vec x}'') \equiv
\frac{B_{\tt s}(\lambda)}{z_0^2} \Big(\frac{kd^2}{8f}\Big)^2    \frac{ 2\mu_0R_{\tt s}}{\alpha\beta} \times\nonumber\\
&&\hskip -55pt \times\,
\frac{ 1}{2\pi} \int_{\phi_-}^{\phi^+} \hskip -8pt d\phi'' \bigg\{
\sqrt{1-\Big(\frac{\rho_0}{r_{\tt s}}\Big)^2\sin^2\phi''}
\bigg(\Big(\frac{2J_1(k d\, \hat u_0(\phi'',\vec x_i))}{k d\, \hat u_0(\phi'',\vec x_i)}\Big)^2+
 \Big(\frac{2J_1(k d\, \hat u_\pi(\phi'',\vec x_i))}{k d\, \hat u_\pi(\phi'',\vec x_i)}\Big)^2\bigg)-\nonumber\\
 &&\hskip -40pt -~
\frac{1}{\alpha r_{\tt s}}\sin\Big[\alpha \rho_0\cos\phi''\Big]\sin\Big[2\alpha r_{\tt s}
\sqrt{1-\Big(\frac{\rho_0}{r_{\tt s}}\Big)^2\sin^2\phi''}\Big]
\Big(\frac{2J_1(k d\, \hat u_0(\phi'',\vec x_i))}{k d\, \hat u_0(\phi'',\vec x_i)}\Big)\Big(\frac{2J_1(k d\, \hat u_\pi(\phi'',\vec x_i))}{k d\, \hat u_\pi(\phi'',\vec x_i)}\Big)\bigg\}.
  \label{eq:P-blur*off4}
\end{eqnarray}

We note that the critical feature of expressions  (\ref{eq:mono-v+})--(\ref{eq:mono-v-}) or (\ref{eq:mono-u-hat-0+})--(\ref{eq:mono-u-hat-pi+}) is the presence of the angular radius of the Einstein deflection angle, $
\theta_{\tt E}$, given by (\ref{eq:alpha-mu}). For a lensing star at a given distance ${\overline z}$ from us, the size of the Einstein ring is determined only by the stellar mass via $r_g$. Furthermore, at the distance ${\overline z}$, the telescope receives light that is characterized by an impact parameter that is typically much larger than  the radius of the lens, $b=\sqrt{2r_g {\overline z}}\gg R_{\tt L}$. This fact justifies our neglecting the contribution from the stellar corona (analog to the solar corona, see \citep{Turyshev-Toth:2020-extend,Turyshev-Toth:2022-broad_SNR,Turyshev-Toth:2022-mono_SNR}) as well as contributions due to the extended nature of the lens (i.e., its multipole moments, see \citep{Turyshev-Toth:2021-multipoles,Turyshev-Toth:2021-imaging,Turyshev-Toth:2021-quartic,Turyshev-Toth:2021-caustics}), effectively treating the stellar gravitational lens as a gravitational monopole.

Results (\ref{eq:FI-go-Int})--(\ref{eq:FI-ir+-Int}) and (\ref{eq:intense2*})--(\ref{eq:P-blur*off4}) represent  the intensity distribution received from an extended source on the image sensor at the focal plane of an imaging telescope. As the telescope moves in the image plane in the focal region of a stellar gravitational lens, the intensity pattern  changes yielding  either of the four distributions. These expressions may be used to model the temporally evolving morphology of an Einstein ring that is formed on the image sensor (e.g., \citep{Cassan-etal:2022}). They can also be used to process imaging data and to recover  parameters of interest regarding the lens and the source.   In Sec.~\ref{sec:m-tools} and \ref{sec:mod-lensevent}, we will discuss the modeling tools that were developed with the help of these expressions and will present imaging simulations of the anticipated lensing events. Using these models and numerical simulations to solve for parameters of interest is yet to be done and is the subject of further study.

\subsection{Power on the image sensor of a telescope}

The image intensities derived in the previous section may now be used to describe the  photometric signals corresponding to the total EM field deposited on the sensor. To study the signal received from a source, we need to evaluate the power received at the telescope's sensor. For that we integrate the intensity over the image sensor over the area taken by the Einstein ring. Similarly to the discussion in \citep{Turyshev-Toth:2020-extend}, it is convenient to estimate this quantity as
{}
\begin{eqnarray}
P_{\tt fp}({\vec x}_0)&=&\epsilon_{\tt ee}P_{\tt }({\vec x}_0)\equiv \epsilon_{\tt ee} \int^{2\pi}_0 \hskip 0pt  d\phi_i \int_0^\infty \hskip -3pt
\rho_i d\rho_i \hskip 0pt I_{\tt }({\vec x}_i,{\vec x}_0),
  \label{eq:pow-frac-bl}
\end{eqnarray}
where $I_{\tt }({\vec x}_i,{\vec x}_0) $ is either from (\ref{eq:intense2*}), (\ref{eq:P-blur*off4}) or from (\ref{eq:FI-go-Int}) and (\ref{eq:FI-ir+-Int}) and  $\epsilon_{\tt ee}\simeq 0.69$, is the encircled energy factor \citep{Turyshev-Toth:2020-extend} describing the fraction of the total energy deposited at the focal plane of the imaging telescope.

To evaluate the integral in (\ref{eq:pow-frac-bl}), we notice that the quantities in the expressions for the intensities  depend on $\rho_i$ and $\phi_i$ via terms that have the form of the type  $(2J_1(x)/x)$. As discussed in \citep{Turyshev-Toth:2020-extend}, for any set of stationery phase solutions, $\phi^{[j]}_\xi, j\in[1,2]$, with $\hat u(\phi^{[j]}_\xi, \vec x_i)$ given by the pairs of (\ref{eq:mono-u-hat-0+})--(\ref{eq:mono-u-hat-pi+}) or (\ref{eq:mono-v+})--(\ref{eq:mono-v-}),  the following two useful relations exist:
{}
\begin{eqnarray}
\int^{2\pi}_0 \hskip -4pt  d\phi_i \int_0^\infty
\hskip -3pt  \rho_i d\rho_i   \Big(\frac{2J_1(k d\, \hat u(\phi^{[j]}_\xi, \vec x_i))}{k d\, \hat u(\phi^{[j]},\vec x_i)}\Big)^2&=& 4\pi \Big(\frac{\lambda f}{\pi d}\Big)^2,
  \label{eq:int-fp1*}\\
\int^{2\pi}_0 \hskip -4pt  d\phi_i \int_0^\infty
\hskip -3pt  \rho_i d\rho_i  \Big(\frac{2J_1(k d\, \hat u(\phi^{[j]}_\xi, \vec x_i))}{k d\, \hat u(\phi^{[j]},\vec x_i)}\Big)\Big(\frac{2J_1(k d\, \hat u(\phi^{[k]}_\xi, \vec x_i))}{k d\, \hat u(\phi^{[k]},\vec x_i)}\Big)&\lesssim&
5\times 10^{-3} \cdot 4\pi \Big(\frac{\lambda f}{\pi d}\Big)^2.
  \label{eq:int-fp2*}
\end{eqnarray}

These expressions can be used to derive the power on the image sensor of a telescope.

\subsubsection{The free-space and weak interference regions}
\label{sec:weak-int2}

Using results (\ref{eq:int-fp1*})--(\ref{eq:int-fp2*}), with the help of (\ref{eq:FI-go-Int}) and (\ref{eq:FI-ir+-Int}), from  (\ref{eq:pow-frac-bl}) we compute the power deposited in the focal plane of a telescope positioned in the free space and in the weak interference regions:
{}
\begin{eqnarray}
P_{\tt fp.0}({\vec x}_0,\lambda)&=&
\epsilon_{\tt ee} \frac{B_{\tt s}(\lambda)}{z_0^2}
 \frac{\pi d^2}{4} \pi R^2_{\tt s},
  \label{eq:fp.pow-go} \\
  P_{\tt fp.weak.int}({\vec x}_0,\lambda)&=&
  \epsilon_{\tt ee} \frac{B_{\tt s}(\lambda)}{z_0^2}
 \frac{\pi d^2}{4} \pi R^2_{\tt s}   \big(a^2_{\tt in}+a^2_{\tt sc}\big)\equiv P_{\tt 0}({\vec x}_0,\lambda) A_{\tt weak.int},
  \label{eq:fp.pow-wi}
\end{eqnarray}
where $A_{\tt weak.int}$ is the microlensing light amplification, that can be evaluated by adding the two amplification factors of the major and minor images from (\ref{eq:a_insc}), which yields the  expression identical to that used in microlensing  \citep{Schneider-Ehlers-Falco:1992,Gaudi:2012}:
  \begin{eqnarray}
A_{\tt weak.int}=a^2_{\tt in}+a^2_{\tt sc}&=&
\frac{1+{4r_g\overline z}/{\rho_0^2}}{\sqrt{1+{8r_g\overline z}/{\rho_0^2}}} \equiv \frac{u^2+2}{u\sqrt{u^2+4}}.
    \label{eq:a12_amp}
\end{eqnarray}
As the microlensing experiments currently deal with unresolved sources, the amplification factor $A_{\tt weak.int}$ has no information on the source and is completely determined by the properties of the lens and the observer's position on the image plane. One can see that the closer the observer is positioned to the primary optical axis, the higher the amplification. Note, however, that this description breaks down as expression (\ref{eq:a12_amp}) diverges as $\rho_0\rightarrow 0$. At small values of $\rho_0$ the amplification factor may be obtained only within the wave-optical treatment, e.g., \citep{Turyshev-Toth:2019-blur,Turyshev-Toth:2019-image,Turyshev-Toth:2020-extend}.

What can be observed in microlensing  is the varying light amplification as the lens moves relative to the source with angular separation between them in units of the Einstein ring, $u={\theta}/{\theta_{\tt E}}$ from (\ref{eq:a_insc}), which may be expressed as 
{}
\begin{equation}
u(t,t_0,u_0,t_{\tt E})=\Big((t-t_0)^2/t^2_{\tt E}+u_0^2\Big)^\frac{1}{2},
\label{eq:u}
\end{equation}
where $t_0$ is the time of closest alignment (which is also the time of maximum amplification in (\ref{eq:a12_amp})), and $u_0$ is the impact parameter of the event, i.e., the angular separation of the source from the lens at $t_0$ expressed  in units of $\theta_{\tt E}$.
The symbol $t_{\tt E}$ represents the characteristic time scale of the event, i.e., the Einstein ring crossing time, $t_{\tt E}=\theta_{\tt E}/\mu_{\rm rel}$, with $\mu_{\rm rel}$ being the relative proper motion between the lens and source; these are the major terms in $u$ that determine the dynamics of the lensing event. Note that, given a particular observing scenario, there may be other terms that should be evaluated and considered, including parallax, observer's motion, various accelerations involved, etc.

What is actually observed during microlensing events is the  change in the flux of the source star that is given as
{}
\begin{eqnarray}
F(t,t_0,u_0,t_{\tt E})=\Big(A_{\tt weak.int}(t,t_0,u_0,t_{\tt E})-1\Big)F_0,
  \label{eq:Amp10}
\end{eqnarray}
where $A_{\tt weak.int}$ is from (\ref{eq:a12_amp}) and $F_0$ is the flux of the unlensed source, which is  not known a priori from the pre-event light curve \citep{Paczynski:1986,Gaudi:2012}.

Eq.~(\ref{eq:Amp10}) immediately reveals the fundamental limitation of microlensing: even if the light curve is observed with perfect accuracy, it yields only four parameters: $t_0, u_0,t_{\tt E}$, and $F_0$. Of these four, only $t_{\tt E}$ contains any information about the lens. Also, it is clear from (\ref{eq:alpha-mu}) and (\ref{eq:Amp10}) that $t_{\tt E}$ has complex dependence on physical quantities of interest.

\subsubsection{The strong interference region}
\label{sec:intense-in-m2}

As discussed in Sec.~\ref{sec:im-form-proc}, temporally resolved imaging offers additional observables that are not available from microlensing. These include the evolving morphology of the phenomenon as the two images of the source move towards their evolving positions along the predetermined circumference of the Einstein ring (e.g., \citep{Turyshev-Toth:2022-wobbles,Cassan-etal:2022}) that eventually forms on the image sensor, widen into arcs, and ultimately (if the source and lens are sufficiently well aligned) join to form that Einstein ring. The relevant photometric information is of interest and should be obtained.

We can now  provide the expression to evaluate the power deposited on the Einstein ring when it is fully formed on the sensor.  For that, using expressions (\ref{eq:intense2*})--(\ref{eq:P-blur*off4}) in (\ref{eq:pow-frac-bl}) and applying results (\ref{eq:int-fp1*})--(\ref{eq:int-fp2*}), we obtain the power received on the image sensor at the focal plane of a telescope at the strong interference region:
  {}
\begin{eqnarray}
P_{\tt fp.ER}({\vec x}_0,\lambda)&=&
\epsilon_{\tt ee} \mu_0\frac{B_{\tt s}(\lambda)}{z_0^2}\frac{\pi d^2 R_{\tt s}}{2\alpha\beta}\eta(\rho_0)
\equiv P_{\tt fp.0}({\vec x}_0,\lambda)A_{\tt ER},
 \label{eq:power0}
\end{eqnarray}
where  following  \citep{Turyshev-Toth:2020-extend,Turyshev-Toth:2019-blur,Turyshev-Toth:2019-image}, we introduced the blur factor due to the spherical aberration of the monopole lens  as
{}
\begin{eqnarray}
\eta(\rho_0)&=&\frac{2}{\pi}\Re \Big\{ {\tt E}\Big[\Big(\frac{\rho_0}{r_{\tt s}}\Big)^2\Big]\Big\},
 \label{eq:power02}
\end{eqnarray}
with ${\tt E}[x]$ being the elliptic integral \citep{Abramovitz-Stegun:1965}  and $\Re\{\}$ is the operation taking the real part of a complex quantity.

The quantity $A_{\tt ER}$ in (\ref{eq:power0}) is the light amplification factor for the extended source as measured at the Einstein ring formed in the strong interference region. Using $\alpha=k\theta_{\tt E}$ from (\ref{eq:alpha}) and $\mu_0=2\pi kr_g$, we obtain $A_{\tt ER}$ as
{}
\begin{eqnarray}
A_{\tt ER}=\frac{P_{\tt fp.ER}}{P_{\tt fp.0}}= \frac{2z_0}{R_{\tt s}}\sqrt{\frac{2r_g}{\overline z}}\eta(\rho_0),
  \label{eq:gain}
\end{eqnarray}
which is independent on the wavelength, as expected \citep{Turyshev-Toth:2020-extend}. This is an analytical expression that can be conveniently and readily evaluated, which provides an accurate representation of light intensity from a source that can be represented by a disk of uniform brightness, amplified by a monopole gravitational lens.

Equations~(\ref{eq:fp.pow-wi}) and (\ref{eq:power0}), describing the power received at different separations from the primary optical axis, $\rho_0$, may now be used to evaluate the  signals in realistic observing conditions. The typical noise sources in these cases are rather small. Indeed, the contribution from the stellar corona is much reduced to the point where it may be neglected.

The amplification factor for extended sources (\ref{eq:gain}) has target-specific and lens-specific terms. Furthermore, for the same distance ${\overline z}$ from the lens, the stellar lenses will provide greater amplification for objects that yield a larger ratio $2z_0/R_{\tt s}$ or  those that are more compact or are located at a greater distance from the lens. Result (\ref{eq:gain})   is always finite and extends (\ref{eq:a12_amp}) in the strong interference region where the distance from the primary optical axis is small. It extends expression (\ref{eq:Amp10}) to cover the entire duration of a lensing event (with $r_{\tt s}=({\overline z}/z_0)R_{\tt s}$):
{}
\begin{eqnarray}
F(t,t_0,u_0,t_{\tt E})=
\begin{cases}
\Big(A_{\tt weak.int}\big(t,t_0,u_0 ,t_{\tt E}\big)-1\Big)F_0, &
~~~~\, \sqrt{\lambda \overline z/\pi}
\lesssim \rho_0 \lesssim \sqrt{2r_g\overline z}-R_{\tt L}+r_{\tt s} \\ \hskip 20pt
A_{\tt ER}\big(t,t_0,u_0,t_{\tt E}, R_{\tt s},z_0\big) F_0, &
~\phantom{\lambda/\theta_{\tt E}+{~~}}0\leq \rho_0 \lesssim \sqrt{\lambda \overline z/\pi} +r_{\tt s}\\
\end{cases},
  \label{eq:Amp10+}
\end{eqnarray}
which is the result that should now be used to model photometric variability of light curves during lensing events. Compared to (\ref{eq:Amp10}), result (\ref{eq:Amp10+}) extends the set of parameters that may be determined during a lensing event, which now includes the size of the source as well as a more direct determination of the lens' mass and distance to the source.

\subsection{Power spectral density in the image plane}
\label{sec:pow-dens}

Results (\ref{eq:fp.pow-go}), (\ref{eq:fp.pow-wi}) and (\ref{eq:power0}) describe the power of the signal received by the telescope at various positions on the image plane and thus at various amplification regimes. These expressions are useful to estimate the signals for a given telescope. It is more convenient to use a telescope agnostic quantity which is the power spectral density deposited on an image plane. To derive the relevant expressions, we divide the power formulae (\ref{eq:fp.pow-go}), (\ref{eq:fp.pow-wi}) and (\ref{eq:power0}) by a factor $\epsilon_{\tt ee}  {\textstyle\frac{1}{4}} \pi d^2$. We also consider imaging a self-luminous resolved source with luminosity $\pi B_{\tt s}(\lambda)=L_{\tt s}(\lambda)/4\pi R_{\tt s}^2$.

As a result, the power spectral density for a direct observation without the stellar gravitational lens, $B_0(\lambda)$, and the same quantity in the weak interference region, $B_{\tt weak.int}(\lambda)$, for the anticipated lensing configurations, where  $\overline z \ll z_0$, have the form (note that because $L_{\tt s}\equiv L_{\tt s}(\lambda)$ these quantities are given in units such as ${\rm W}/{{\rm m}^2}/{\rm nm}$):
{}
\begin{eqnarray}
B_{\tt 0}(\lambda)=
\frac{L_{\tt s}}{4\pi z_0^2},
\qquad
B_{\tt weak.int}(\lambda)=
\frac{L_{\tt s}}{4\pi z_0^2}  \frac{u^2+2}{u\sqrt{u^2+4}}
\equiv B_{\tt 0}(\lambda) A_{\tt weak.int},
  \label{eq:Pexo-pdsl0}
\end{eqnarray}
where $A_{\tt weak.int}$ is from (\ref{eq:a12_amp}).
The same quantity in the image plane in the strong interference region is given as
{}
\begin{eqnarray}
B_{\tt ER}(\lambda)= \frac{ L_{\tt s}}{4\pi R_{\tt s}^2} \frac{2R_{\tt s}}{z_0}\sqrt{\frac{2r_g}{\overline z}}\eta(\rho_0)
\equiv B_{\tt 0}(\lambda) A_{\tt ER},
  \label{eq:Pexo-pdsl}
\end{eqnarray}
where $\eta(\rho_0)$ is from (\ref{eq:power0}) and $R_{\tt s}$ is the source's radius and $z_0$ is its distance from the lens.

Is the amplification (\ref{eq:gain}) enough to see the ring in the presence of the  lensing star's light? To evaluate the contrast ratio, we take the ratio of the Einstein ring's power density (\ref{eq:Pexo-pdsl}) and that of the lens (\ref{eq:Pexo-pdsl0}):
{}
\begin{eqnarray}
{\tt C}_{{\tt ER}}=\frac{B_{\tt ER}}{B_{\tt 0L}}= \frac{ L_{\tt s}}{4\pi R_{\tt s}^2} \frac{2R_{\tt s}}{z_0}\sqrt{\frac{2r_g}{\overline z}}\eta(\rho_0)\Big/\frac{L_{\tt L}}{4\pi {\overline z}^2} =\Big(\frac{ L_{\tt s}}{L_{\tt L}}\Big) \Big(\frac{\overline z}{R_{\tt s}}\Big)\frac{2 \sqrt{{2r_g\overline z}}}{z_0}\eta(\rho_0).
  \label{eq:gain_L}
\end{eqnarray}

The set of expressions derived above may now be used to estimate realistic signal levels that may be expected in the lensing events involving stars in our neighborhood, which will be done in Sec.~\ref{sec:ER-image} below.

\subsection{Science with resolved imaging during the encounter}
\label{sec:im-sci}

Our previous work on the formation of the Einstein rings done in the context of our study of the solar gravitational lens (e.g., \citep{Turyshev-Toth:2022-wobbles}) informs us on the prospective imaging observations with generic stellar lenses. In particular, temporally resolved, high-contrast imaging of the formation of an Einstein ring around a nearby star provides significant new information about both the lens and the lensed background object. What distinguishes such observations from more common microlensing events that consider only photometric flux variability? The different stages of forming a resolved Einstein ring offer the following observables that complement conventional microlensing observations:
{}
\begin{itemize}
\item The temporally evolving measurements of the shape, size and orientation of the primary and secondary
images as before and after the formation of the Einstein ring;
\item Minute variations of brightness alongside the ring's circumference;
\item Direct measurements of the ring brightness (as opposed to photometric measurements of the combined brightness of ring and lens);
\item Direct spectroscopy measurements of the ring, which are relatively uncontaminated by light from the lens;
\item Sky positions of secondary lensing events, such as temporary brightening due to lensing by a planet orbiting the lens star.
\end{itemize}

Such observations can be used in conjunction with microlensing photometric observables to deepen our understanding of the temporal evolution of the lensing event, its three-dimensional geometry, and the physical properties of both the lens and the lensed object.  Most of that insight comes from studying our investigations of the intensity distribution on the telescope's image sensor (i.e., images) that are given by (\ref{eq:FI-ir+-Int}) and (\ref{eq:intense2*})--(\ref{eq:P-blur*off4}). Specifically:
{}
\begin{itemize}
\item The angular size of the emerging Einstein ring, $\theta_{\tt E}$, provides an independent estimate of the mass of the lens, as seen in (\ref{eq:alpha-mu}).
\item The orientation of the primary and secondary images of the source and subsequently, the emerging incomplete arcs of the Einstein ring offer precise information about the direction of motion of the source  with respect to the lens, as given by the normalized spatial frequencies (\ref{eq:mono-u-hat-0+})--(\ref{eq:mono-u-hat-pi+}) and discussed in \citep{Turyshev-Toth:2022-wobbles}. Together with the temporal evolution of the event and the velocity-dependent Doppler shift of the lensed object's spectrum (comes from the spectral content of $L_{\tt s}(\lambda)$, see \citep{Turyshev-Toth:2022-mono_SNR}), a three-dimensional reconstruction of the encounter may be possible. In addition, there is a possibility to determine the rotational state of the source via spectroscopic measurements.
\item Variations of brightness along the Einstein ring's circumference may offer clues about the source's structure.
\item As the extremely narrow field of view (FOV) of the lens traverses the lensed object, small temporal variations in the ring's overall brightness map the surface brightness of the source alongside a narrow cross-sectional track. If several telescopes at various positions on the image plane are involved in observing the lensing event by taking time-series of photometric data,  one can attempt to reconstruct the image of the source (as was demonstrated in the context of our study of the solar gravitational lens \citep{Turyshev-Toth:2019-extend,Turyshev-Toth:2020-extend,Toth-Turyshev:2020}).
\item Spectroscopy of the Einstein ring, relatively uncontaminated by light from the lensing star, may offer a better understanding of the physical nature of the source (as was shown in \citep{Turyshev-Toth:2022-mono_SNR}.)
\item Brief brightening of the primary or secondary images may indicate the presence of an exoplanet orbiting the lens.   The evolving ring's morphology may allow one to recover the information on the mass of the planet and its orbital parameters.
\item If multiple lenses with a similar luminous background are available, this may allow for a broader investigations on the matter distribution within that background.
\end{itemize}

As these examples show, the additional information provided through the size, temporal evolution, orientation, brightness, spectroscopy  and other details of the evolving morphology of the emerging Einstein ring adds a significant new dimension to lensing events. We have not yet endeavored to quantify potential results, but it is clear that once specific lensing targets and instruments are considered, these new observables offer very valuable constraints on both the lensing system and source in the form of information that is not available through microlensing alone and cannot be obtained by other means.

\section{Gravitational lenses in our stellar neighborhood}
\label{sec:ER-image}

The advantage of considering stars in our stellar neighborhood as lenses is that the resulting Einstein ring has a large angular size, well separated from the star and is resolvable by modern instruments. Consider a lensing star with radius $R_{\tt L}$, the ratio of its angular radius $\theta_{ 0}=R_{\tt L}/\overline z$ to the angular radius of the Einstein ring is estimated to be $\theta_{\tt E}/\theta_{ 0}=\sqrt{2r_g \overline z}/R_{\tt L}=61.37 ({M_{\tt L}}/{M_\odot})^\frac{1}{2}(\overline z/10~{\rm pc})^\frac{1}{2}(R_\odot/R_{\tt L})$. The angular features of this size may be  detectable in many cases of interest. Furthermore, the Einstein ring formed at such a large separation from the star correspond to the light ray with impact parameters of $b=\theta_{\tt E} \overline z=\sqrt{2r_g \overline z}=0.295({M_{\tt L}}/{M_\odot})^\frac{1}{2}(\overline z/10~{\rm pc})^\frac{1}{2}$~AU. At such separations from the lens, the lensing  is not sensitive to the mass distribution inside the lensing star, thus there is no need to consider multipole moments \citep{Turyshev-Toth:2021-multipoles}. In addition, the stellar corona contribution on the light propagation may also be neglected and the lens is fully described by the PSF of a gravitational monopole  \citep{Turyshev-Toth:2017,Turyshev-Toth:2020-extend}.

\subsection{Power density estimates for realistic sources}

\begin{table*}[t!]
\caption{\label{tb:pdreceive2}
Physical parameters including distance, mass, size and temperature, and derived values for 50 nearby less luminous stars and brown dwarfs, and 5 type G stars, along with the angular size and brightness of the Einstein ring that would form from light emitted by a Sun-like star (i.e., $M_{\tt s}=M_\odot$ and $R_{\tt s}=R_\odot$) located at $z_0=8$~kpc directly behind the lens; also $\theta_0=R_{\tt L}/{\overline z}$. The Einstein ring's apparent angular radius $\theta_{\tt E}$ may be compared against the diffraction-limited resolution of a $d=10$~m aperture telescope at $\lambda=0.6~\mu$m, $\theta=1.22\lambda/d\sim 15$~mas. Bolometric luminosities ($L$) are calculated from temperature and radii using the Stefan--Boltzmann law. The apparent bolometric magnitude $m_{\tt bolo}$ and the magnitude $m_{\tt ER}$ of the Einstein ring are calculated without regard for extinction. Number of known planets and yet to be confirmed planetary candidates is also shown.}\scriptsize\hskip -1.25in
\begin{tabular}{|l|c|c|c|c||c|c|c|c|c|c|c|c|c|}\hline
                       &${\overline z}$ &$M_{\tt L}/M_\odot$&$R_{\tt L}/R_\odot$& $T$ &$\theta_0$&$\theta_{\tt E}$        &$L_{\tt L}/L_\odot$  &  $B_{\tt 0L}$  &  $B_{\tt ER}$               &${\tt C}_{\tt ER}$   &$m_{\rm bolo}$&$ m_{\tt ER}$&Exo-\\
Name                   &  pc            &             &             &  K  & \multicolumn{2}{c|}{mas}          &             &\multicolumn{2}{c|}{ $10^{-9}$~W/m$^2$ }      &               &              &             & planets\\
\hline\hline
\multicolumn{14}{|l|}{\bf Nearby dim stars and brown dwarfs\protect\footnote{\protect\url{https://en.wikipedia.org/wiki/List_of_nearest_stars_and_brown_dwarfs}.}}\\\hline
Proxima Centauri       & 1.3\phantom{0} & 0.122\phantom{} & 0.154\phantom{} & \phantom{0}3042 & 0.55 & 27.65  & 0.0018\phantom{00} & \phantom{0}0.035\phantom{000} & 0.048 & \phantom{000}1.37 & \phantom{0}7.16 & 6.81 & 2 + 1?\\
Barnard's Star         & 1.83\phantom{} & 0.144\phantom{} & 0.196\phantom{} & \phantom{0}3134 & 0.50 & 25.32  & 0.0033\phantom{00} & \phantom{0}0.032\phantom{000} & 0.044 & \phantom{000}1.37 & \phantom{0}7.25 & 6.90 & 1?\\
Luhman A               & 1.99\phantom{} & 0.032\phantom{} & 0.085\phantom{} & \phantom{0}1350 & 0.20 & 11.43  & 0.000022\phantom{} & \phantom{0}0.00017\phantom{0} & 0.020 & \phantom{0}113.34 & \phantom{}12.91 & 7.77 & \\
Luhman B               & 1.99\phantom{} & 0.027\phantom{} & 0.104\phantom{} & \phantom{0}1210 & 0.24 & 10.50  & 0.000021\phantom{} & \phantom{0}0.00017\phantom{0} & 0.018 & \phantom{0}107.76 & \phantom{}12.94 & 7.86 & \\
Wolf 359               & 2.41\phantom{} & 0.11\phantom{0} & 0.144\phantom{} & \phantom{0}2749 & 0.28 & 19.28  & 0.0011\phantom{00} & \phantom{0}0.0059\phantom{00} & 0.033 & \phantom{000}5.66 & \phantom{0}9.08 & 7.20 & 2\\
Lalande 21185          & 2.55\phantom{} & 0.39\phantom{0} & 0.392\phantom{} & \phantom{0}3601 & 0.72 & 35.32  & 0.023\phantom{000} & \phantom{0}0.115\phantom{000} & 0.061 & \phantom{000}0.53 & \phantom{0}5.86 & 6.54 & 2\\
Sirius B               & 2.64\phantom{} & 1.018\phantom{} & 0.008\phantom{} & \phantom{}25000 & 0.01 & 56.04  & 0.025\phantom{000} & \phantom{0}0.114\phantom{000} & 0.097 & \phantom{000}0.85 & \phantom{0}5.86 & 6.04 & \\
Luyten 726-8 A         & 2.72\phantom{} & 0.102\phantom{} & 0.14\phantom{0} & \phantom{0}2670 & 0.24 & 17.48  & 0.00089\phantom{0} & \phantom{0}0.0039\phantom{00} & 0.030 & \phantom{000}7.77 & \phantom{0}9.53 & 7.31 & \\
Luyten 726-8 B         & 2.68\phantom{} & 0.1\phantom{00} & 0.14\phantom{0} & \phantom{0}2650 & 0.24 & 17.45  & 0.00087\phantom{0} & \phantom{0}0.0039\phantom{00} & 0.030 & \phantom{000}7.73 & \phantom{0}9.53 & 7.31 & \\
Ross 154               & 2.98\phantom{} & 0.17\phantom{0} & 0.24\phantom{0} & \phantom{0}3340 & 0.38 & 21.57  & 0.0064\phantom{00} & \phantom{0}0.023\phantom{000} & 0.037 & \phantom{000}1.59 & \phantom{0}7.59 & 7.08 & \\
Ross 248               & 3.16\phantom{} & 0.136\phantom{} & 0.16\phantom{0} & \phantom{0}2799 & 0.24 & 18.72  & 0.0014\phantom{00} & \phantom{0}0.0045\phantom{00} & 0.032 & \phantom{000}7.12 & \phantom{0}9.37 & 7.23 & \\
Lacaille 9352          & 3.29\phantom{} & 0.486\phantom{} & 0.47\phantom{0} & \phantom{0}3692 & 0.67 & 34.70  & 0.037\phantom{000} & \phantom{0}0.110\phantom{000} & 0.060 & \phantom{000}0.55 & \phantom{0}5.91 & 6.56 & 2 + 1?\\
Ross 128               & 3.37\phantom{} & 0.168\phantom{} & 0.197\phantom{} & \phantom{0}3192 & 0.27 & 20.13  & 0.0036\phantom{00} & \phantom{0}0.010\phantom{000} & 0.035 & \phantom{000}3.42 & \phantom{0}8.49 & 7.15 & 1\\
Alpha Canis Minoris B  & 3.51\phantom{} & 0.602\phantom{} & 0.012\phantom{} & \phantom{0}7740 & 0.02 & 37.37  & 0.00049\phantom{0} & \phantom{0}0.0013\phantom{00} & 0.065 & \phantom{00}50.41 & \phantom{}10.74 & 6.48 & \\
Struve 2398 A          & 3.52\phantom{} & 0.334\phantom{} & 0.351\phantom{} & \phantom{0}3441 & 0.46 & 27.79  & 0.015\phantom{000} & \phantom{0}0.040\phantom{000} & 0.048 & \phantom{000}1.19 & \phantom{0}7.00 & 6.80 & \\
Struve 2398 B          & 3.52\phantom{} & 0.248\phantom{} & 0.273\phantom{} & \phantom{0}3345 & 0.36 & 23.94  & 0.0084\phantom{00} & \phantom{0}0.022\phantom{000} & 0.041 & \phantom{000}1.91 & \phantom{0}7.67 & 6.96 & 2\\
Groombridge 34 A       & 3.56\phantom{} & 0.38\phantom{0} & 0.38\phantom{0} & \phantom{0}3607 & 0.50 & 29.47  & 0.022\phantom{000} & \phantom{0}0.056\phantom{000} & 0.051 & \phantom{000}0.92 & \phantom{0}6.65 & 6.74 & 2\\
Groombridge 34 B       & 3.56\phantom{} & 0.15\phantom{0} & 0.18\phantom{0} & \phantom{0}3304 & 0.24 & 18.52  & 0.0035\phantom{00} & \phantom{0}0.0088\phantom{00} & 0.032 & \phantom{000}3.64 & \phantom{0}8.65 & 7.24 & \\
DX Cancri              & 3.58\phantom{} & 0.09\phantom{0} & 0.11\phantom{0} & \phantom{0}2840 & 0.14 & 14.31  & 0.00071\phantom{0} & \phantom{0}0.0018\phantom{00} & 0.025 & \phantom{00}13.95 & \phantom{}10.39 & 7.52 & \\
Gliese 1061            & 3.67\phantom{} & 0.113\phantom{} & 0.156\phantom{} & \phantom{0}2999 & 0.20 & 15.83  & 0.0018\phantom{00} & \phantom{0}0.0042\phantom{00} & 0.027 & \phantom{000}6.49 & \phantom{0}9.45 & 7.41 & 3\\
YZ Ceti                & 3.72\phantom{} & 0.13\phantom{0} & 0.168\phantom{} & \phantom{0}3056 & 0.21 & 16.88  & 0.0022\phantom{00} & \phantom{0}0.0051\phantom{00} & 0.029 & \phantom{000}5.67 & \phantom{0}9.23 & 7.34 & 3 + 1?\\
Luyten's Star          & 3.79\phantom{} & 0.26\phantom{0} & 0.35\phantom{0} & \phantom{0}3150 & 0.43 & 23.65  & 0.011\phantom{000} & \phantom{0}0.024\phantom{000} & 0.041 & \phantom{000}1.68 & \phantom{0}7.55 & 6.98 & 2 + 2?\\
Teegarden's Star       & 3.80\phantom{} & 0.089\phantom{} & 0.107\phantom{} & \phantom{0}2904 & 0.13 & 13.75  & 0.00073\phantom{0} & \phantom{0}0.0016\phantom{00} & 0.024 & \phantom{00}14.84 & \phantom{}10.50 & 7.57 & 2\\
Kapteyn's Star         & 3.93\phantom{} & 0.281\phantom{} & 0.291\phantom{} & \phantom{0}3570 & 0.34 & 24.12  & 0.012\phantom{000} & \phantom{0}0.026\phantom{000} & 0.042 & \phantom{000}1.62 & \phantom{0}7.49 & 6.96 & 2?\\
Lacaille 8760          & 3.97\phantom{} & 0.6\phantom{00} & 0.51\phantom{0} & \phantom{0}3800 & 0.60 & 35.09  & 0.049\phantom{000} & \phantom{0}0.099\phantom{000} & 0.061 & \phantom{000}0.61 & \phantom{0}6.02 & 6.55 & \\
SCR 1845-6357 A        & 4.00\phantom{} & 0.07\phantom{0} & 0.096\phantom{} & \phantom{0}2700 & 0.11 & 11.93  & 0.00044\phantom{0} & \phantom{0}0.00088\phantom{0} & 0.021 & \phantom{00}23.38 & \phantom{}11.15 & 7.72 & \\
Kruger 60 A            & 4.01\phantom{} & 0.271\phantom{} & 0.35\phantom{0} & \phantom{0}3180 & 0.41 & 23.46  & 0.011\phantom{000} & \phantom{0}0.022\phantom{000} & 0.041 & \phantom{000}1.80 & \phantom{0}7.63 & 6.99 & \\
Kruger 60 B            & 4\phantom{.00} & 0.176\phantom{} & 0.24\phantom{0} & \phantom{0}2890 & 0.28 & 18.93  & 0.0036\phantom{00} & \phantom{0}0.0073\phantom{00} & 0.033 & \phantom{000}4.51 & \phantom{0}8.86 & 7.22 & \\
DEN 1048-3956          & 4.05\phantom{} & 0.075\phantom{} & 0.108\phantom{} & \phantom{0}2330 & 0.12 & 12.29  & 0.00031\phantom{0} & \phantom{0}0.00061\phantom{0} & 0.021 & \phantom{00}35.00 & \phantom{}11.55 & 7.69 & \\
Wolf 1061              & 4.31\phantom{} & 0.294\phantom{} & 0.307\phantom{} & \phantom{0}3342 & 0.33 & 23.58  & 0.011\phantom{000} & \phantom{0}0.018\phantom{000} & 0.041 & \phantom{000}2.23 & \phantom{0}7.85 & 6.98 & 3\\
Van Maanen's Star      & 4.31\phantom{} & 0.67\phantom{0} & 0.014\phantom{} & \phantom{0}6130 & 0.01 & 35.56  & 0.00024\phantom{0} & \phantom{0}0.00042\phantom{0} & 0.061 & \phantom{0}147.28 & \phantom{}11.96 & 6.53 & 1?\\
Gliese 1               & 4.35\phantom{} & 0.33\phantom{0} & 0.33\phantom{0} & \phantom{0}3478 & 0.35 & 24.87  & 0.014\phantom{000} & \phantom{0}0.024\phantom{000} & 0.043 & \phantom{000}1.76 & \phantom{0}7.54 & 6.92 & \\
L 1159-16              & 4.47\phantom{} & 0.14\phantom{0} & 0.161\phantom{} & \phantom{0}3158 & 0.17 & 15.97  & 0.0023\phantom{00} & \phantom{0}0.0037\phantom{00} & 0.028 & \phantom{000}7.41 & \phantom{0}9.58 & 7.40 & 2 + 1?\\
Gliese 687             & 4.55\phantom{} & 0.401\phantom{} & 0.492\phantom{} & \phantom{0}3095 & 0.50 & 26.79  & 0.020\phantom{000} & \phantom{0}0.031\phantom{000} & 0.046 & \phantom{000}1.49 & \phantom{0}7.28 & 6.84 & 2\\
Gliese 674             & 4.55\phantom{} & 0.353\phantom{} & 0.361\phantom{} & \phantom{0}3404 & 0.37 & 25.13  & 0.016\phantom{000} & \phantom{0}0.024\phantom{000} & 0.043 & \phantom{000}1.78 & \phantom{0}7.54 & 6.91 & 1\\
LHS 292                & 4.56\phantom{} & 0.08\phantom{0} & 0.11\phantom{0} & \phantom{0}2900 & 0.11 & 11.95  & 0.00077\phantom{0} & \phantom{0}0.0012\phantom{00} & 0.021 & \phantom{00}17.37 & \phantom{}10.82 & 7.72 & \\
LP 145-141             & 4.64\phantom{} & 0.75\phantom{0} & 0.01\phantom{0} & \phantom{0}8500 & 0.01 & 36.30  & 0.00047\phantom{0} & \phantom{0}0.00070\phantom{0} & 0.063 & \phantom{00}89.43 & \phantom{}11.39 & 6.51 & \\
Gliese 876             & 4.67\phantom{} & 0.37\phantom{0} & 0.376\phantom{} & \phantom{0}3129 & 0.37 & 25.40  & 0.012\phantom{000} & \phantom{0}0.018\phantom{000} & 0.044 & \phantom{000}2.45 & \phantom{0}7.87 & 6.90 & 4 + 2?\\
Gliese 412 A           & 4.90\phantom{} & 0.48\phantom{0} & 0.38\phantom{0} & \phantom{0}3687 & 0.36 & 28.23  & 0.024\phantom{000} & \phantom{0}0.032\phantom{000} & 0.049 & \phantom{000}1.52 & \phantom{0}7.25 & 6.79 & \\
Gliese 412 B           & 4.91\phantom{} & 0.1\phantom{00} & 0.13\phantom{0} & \phantom{0}2700 & 0.12 & 12.88  & 0.00081\phantom{0} & \phantom{0}0.0011\phantom{00} & 0.022 & \phantom{00}20.66 & \phantom{}10.93 & 7.64 & \\
AD Leonis              & 4.97\phantom{} & 0.405\phantom{} & 0.39\phantom{0} & \phantom{0}3390 & 0.37 & 25.77  & 0.018\phantom{000} & \phantom{0}0.024\phantom{000} & 0.045 & \phantom{000}1.89 & \phantom{0}7.58 & 6.88 & \\
Gliese 832             & 4.97\phantom{} & 0.45\phantom{0} & 0.48\phantom{0} & \phantom{0}3620 & 0.45 & 27.16  & 0.035\phantom{000} & \phantom{0}0.046\phantom{000} & 0.047 & \phantom{000}1.01 & \phantom{0}6.84 & 6.83 & 2\\
Gliese 682             & 5.01\phantom{} & 0.27\phantom{0} & 0.3\phantom{00} & \phantom{0}3237 & 0.28 & 20.95  & 0.0089\phantom{00} & \phantom{0}0.011\phantom{000} & 0.036 & \phantom{000}3.18 & \phantom{0}8.37 & 7.11 & 2?\\
Omicron Eridani A      & 5.01\phantom{} & 0.84\phantom{0} & 0.81\phantom{0} & \phantom{0}5300 & 0.75 & 36.95  & 0.464\phantom{000} & \phantom{0}0.596\phantom{000} & 0.064 & \phantom{000}0.11 & \phantom{0}4.07 & 6.49 & \\
Omicron Eridani B      & 5.01\phantom{} & 0.573\phantom{} & 0.014\phantom{} & \phantom{}16500 & 0.01 & 30.53  & 0.013\phantom{000} & \phantom{0}0.017\phantom{000} & 0.053 & \phantom{000}3.15 & \phantom{0}7.95 & 6.70 & 1\\
Omicron Eridani C      & 5.01\phantom{} & 0.204\phantom{} & 0.31\phantom{0} & \phantom{0}3100 & 0.29 & 18.19  & 0.0080\phantom{00} & \phantom{0}0.010\phantom{000} & 0.031 & \phantom{000}3.08 & \phantom{0}8.49 & 7.26 & \\
EV Lacertae            & 5.05\phantom{} & 0.35\phantom{0} & 0.36\phantom{0} & \phantom{0}3400 & 0.33 & 23.75  & 0.016\phantom{000} & \phantom{0}0.020\phantom{000} & 0.041 & \phantom{000}2.09 & \phantom{0}7.78 & 6.97 & \\
GJ 3379                & 5.21\phantom{} & 0.231\phantom{} & 0.246\phantom{} & \phantom{0}3284 & 0.22 & 19.01  & 0.0063\phantom{00} & \phantom{0}0.0075\phantom{00} & 0.033 & \phantom{000}4.39 & \phantom{0}8.83 & 7.21 & \\
DENIS J081730.0-615520 & 5.21\phantom{} & 0.015\phantom{} & 0.094\phantom{} & \phantom{00}950 & 0.08 & ~4.84  & 0.000006\phantom{} & \phantom{0}0.000008\phantom{} & 0.008 & \phantom{}1092.56 & \phantom{}16.30 & 8.70 & \\
Gliese 445             & 5.25\phantom{} & 0.14\phantom{0} & 0.285\phantom{} & \phantom{0}3507 & 0.25 & 14.73  & 0.011\phantom{000} & \phantom{0}0.013\phantom{000} & 0.025 & \phantom{000}1.98 & \phantom{0}8.24 & 7.49 & \\
\hline\hline
\multicolumn{14}{|l|}{\bf Nearby type G stars}\\\hline
Alpha Centauri A       & 1.34\phantom{} & 1.079\phantom{} & 1.218\phantom{} & \phantom{0}5790 & 4.23 & 80.97  & 1.495\phantom{000} & \phantom{}26.69\phantom{0000} & 0.140 & \phantom{000}0.01 &\phantom{}--0.06 & 5.64 & 1?\\
Tau Ceti               & 3.65\phantom{} & 0.783\phantom{} & 0.793\phantom{} & \phantom{0}5344 & 1.01 & 41.79  & 0.460\phantom{000} & \phantom{0}1.110\phantom{000} & 0.072 & \phantom{000}0.07 & \phantom{0}3.40 & 6.36 & 4 + 5?\\
Eta Cassiopeia A       & 5.92\phantom{} & 0.972\phantom{} & 1.039\phantom{} & \phantom{0}5973 & 0.82 & 36.56  & 1.232\phantom{000} & \phantom{0}1.130\phantom{000} & 0.063 & \phantom{000}0.06 & \phantom{0}3.38 & 6.50 & \\
82 G. Eridani          & 6.04\phantom{} & 0.7\phantom{00} & 0.92\phantom{0} & \phantom{0}5401 & 0.71 & 30.72  & 0.646\phantom{000} & \phantom{0}0.570\phantom{000} & 0.053 & \phantom{000}0.09 & \phantom{0}4.12 & 6.69 & 3 + 3?\\
Delta Pavonis          & 6.10\phantom{} & 1.051\phantom{} & 1.197\phantom{} & \phantom{0}5571 & 0.91 & 37.46  & 1.238\phantom{000} & \phantom{0}1.071\phantom{000} & 0.065 & \phantom{000}0.06 & \phantom{0}3.43 & 6.48 & 1?\\\hline
\end{tabular}
\end{table*}

Using the expression presented in Sec.~\ref{sec:pow-dens}, we can estimate the photon power density that may be anticipated from a background star in the image plane positioned in the strong interference region of the stellar gravitational lens as viewed from within our solar system (similarly to the approach used in \citep{Turyshev-Toth:2020-extend,Turyshev-Toth:2022-wobbles}). Clearly, ideal lensing conditions are offered by stars that are relatively near the solar system (so that the angular radius of the resulting Einstein ring is large) and are less luminous (yielding a higher contrast ratio, $\tt C_{ER}$, between Einstein ring and lens). Even with these restrictions, there are plenty of potential lenses in the solar system's neighborhood.

Some potential lens candidates are shown in Table~\ref{tb:pdreceive2}. This table lists 50 less luminous stars in the solar system's immediate vicinity, along with 5 type G stars that are brighter but more closely resemble the Sun. In addition to their physical properties, the calculated angular diameter and brightness of the Einstein ring is also shown, assuming conjunction with a light source that is a Sun-like star at $z_0=8$~kpc from the solar system.

Many more stars beyond those listed in this table may act as lenses with resolvable Einstein rings. As the angular size of the Einstein ring varies as the inverse square root of distance, stars up to $\sim$100 pc may offer opportunities to go beyond photometric microlensing observations and study the geometry and evolving morphology of the ring.

It is interesting that the stellar lensing phenomena may present both an opportunity and a challenge.  The Einstein ring can reveal information about both the lensing star and the objects being lensed.  However, the gravitationally amplified light from distant background objects can contaminate the lensing observations if their position is very close to the primary optical axis and thus their image may coincide with the Einstein ring that is formed around the lens.  For instance, from Table \ref{tb:pdreceive2}, we see that the location of the Einstein ring around Proxima Cen A is $\theta_{\tt E}\overline z=0.036$~AU, which is very close to the orbit  of Proxima b, the exoplanet that was recently discovered in that system with a semi-major axis of $0.04856\pm0.00030$~AU \citep{Faria-etal:2022}. It may be important to look for other, similar coincidences and take them into account when planning for the direct detection of exoplanets around nearby stars.

We note that when the contrast ratio ${\tt C_{ER}}$, given by (\ref{eq:gain_L}), is high, the Einstein ring may be observable even against the glare of the lens (see Sec.~\ref{sec:mod-lensevent} and figure  Fig.~\ref{fig:lensing} therein for details and a simulated view). In other cases, the contrast ratio can be improved by using a coronagraph, similar to those used for exoplanet search \citep{Traub-Oppenheimer:2010}. For optimal observation, coronagraphs require the IWA \citep{CarrionGonzalez-etal:2021} to be better than $3\lambda/d$; considering the sizes of the Einstein rings for the nearby G-type stars from Table~\ref{tb:pdreceive2} to be $\theta_{\tt E}^{\tt G}\gtrsim35$~ mas, implies that resolving the Einstein ring with a high contrast ratio would require a telescope with aperture $d\simeq3\lambda/\theta^{\tt G}_{\tt E}=10.6 ~(\lambda/0.6~\mu{\rm m})$ m. Such monolithic apertures with a diffraction-limited performance may potentially be realized in space, unaffected by atmospheric blur, notably, with the upcoming capabilities of the Starship launch vehicle.\footnote{\url{https://www.spacex.com/vehicles/starship/}}

\begin{table*}[t]
\vskip-15pt
\caption{Capabilities of the some of the current and near future optical telescopes. (Note that future dates are likely to change.)  Angular resolution is at the shortest wavelength, as reported by the projects.  Both ground and space-based instruments are presented. Several of these optical telescopes have coronagraphic instruments that may help with observing Einstein rings around nearby stars.
\label{tb:tel-cap}}
\begin{tabular}{|l|c|c|c|c|}\hline
Telescope  & Aperture & First light & Bandwidth &  Resolution \\
           &     m    &             &  $\mu$m   &     mas     \\\hline\hline
\multicolumn{5}{|l|}{\bf Ground-based}\\\hline
Keck  I/II (and interferometer) & 10.0  & 1990,~1996& 0.3--5.0 & 22\footnote{\citep{Wizinowich_2000}} (5\footnote{\url{https://www2.keck.hawaii.edu/ASTRA/documents/intro.pdf}}) \\
Very Large Telescope (VLT) (and interferometer) & 8.2  & 1998--2000& 0.3--20 & 50 (2)\footnote{\url{https://www.eso.org/public/about-eso/faq/faq-vlt-paranal/\#1}} \\
Subaru Telescope & 8.2  & 1999& 0.3--30 & 200\footnote{\url{https://www.subarutelescope.org/old/Introduction/telescope.html}} \\
Southern African Large Telescope (SALT) & $11.1 \times 9.8$ & 2005& 0.32--1.7 & 600\footnote{\citep{SALT1995}} \\
Vera C. Rubin Observatory & 8.36  & 2023& 0.32--1.06 & 200\footnote{\url{https://www.lsst.org/scientists/keynumbers}} \\
Thirty Meter Telescope  (TMT) & 30.0  & 2027 & 	0.31--28 & 7\footnote{\citep{TMT2016}} \\
Extremely Large Telescope (ELT) & 39.3  & 2027& 0.47--19.0 & 10\footnote{\citep{2010SPIE.7735E..2EK,2021Msngr.182....7T}} \\
Giant Magellan Telescope (GMT) & 24.5  & 2029&  0.32--25 & 10\footnote{\url{https://en.wikipedia.org/wiki/Giant_Magellan_Telescope}} \\
\hline\hline
\multicolumn{5}{|l|}{\bf In space }\\\hline
Hubble Space Telescope  (HST) & 2.4   & 1990& 0.2--2.4 & 40\footnote{\url{https://www.nasa.gov/missions/highlights/webcasts/shuttle/sts109/hubble-qa.html}} \\
James Webb Space Telescope  (JWST) & 6.5  & 2022& 	0.6--28.3 &  30\footnote{\url{https://jwst-docs.stsci.edu/jwst-near-infrared-camera/nircam-performance/nircam-point-spread-functions}} \\
Nancy Grace Roman Space Telescope & 2.4  & 2027& 0.48--2.30 & 110\footnote{\citep{2017IAUS..325..373H}} \\
\hline
\end{tabular}
\end{table*}

\subsection{Observational capabilities}

Can existing astronomical instruments or instruments currently under development image Einstein rings that appear around stars in the vicinity of the solar system? Considering the published capabilities of leading edge observatories, the answer appears affirmative. Optical telescopes currently in operation, including
Subaru\footnote{\url{https://subarutelescope.org/en/}},
Keck I/II\footnote{\url{https://www.keckobservatory.org/}},
the Hubble Space Telescope (HST)\footnote{\url{https://www.stsci.edu/hst}},
and the James Webb Space Telescope (JWST)\footnote{\url{https://webbtelescope.org/}},
offer good capabilities to observe Einstein rings that may form around some nearby stars. With the new generation of telescopes, such as the
Extremely Large Telescope (ELT)\footnote{\url{https://elt.eso.org/}},
the Thirty Meter Telescope (TMT)\footnote{\url{https://www.tmt.org/}},
or the Vera C. Rubin Observatory (formerly LSST)\footnote{\url{https://www.lsst.org/}},
this situation is expected to further improve. Therefore, it is time to reassess modern capabilities in comparison with what is needed to image stellar gravitational lensing events, to see what may be possible today and what will become available in the near future.

Considering existing and planned observational capabilities, we note that there are several instruments that potentially could offer useful performance. However, as shown in Table~\ref{tb:tel-cap}, the angular resolution of these instruments is usually below their theoretical diffraction limit. Terrestrial telescopes, in particular, have to contend with atmospheric blur, and run into limits even when adaptive optics are used. Other factors that influenced telescope or instrument design may play a role: simply, the telescope might not have been optimized for maximum angular resolution at the shortest observable wavelength.

Considering the theoretical full width at half maximum (FWHM) of these telescopes, $\theta\sim \lambda/d$, most of  the instruments in Table~\ref{tb:tel-cap} should have the capability to resolve at least some of the Einstein rings listed in Table~\ref{tb:pdreceive2}. However, reality is quite different. Most of the ground-based instruments have segmented mirrors that also operate through the atmospheric turbulence which degrade their capabilities. Space-based telescopes either have modest apertures (i.e., HST, Nancy Roman) or else have segmented mirrors (i.e., JWST), that impact their PSF and maximum resolution. However, most of the new space telescopes may have coronographic instruments (i.e., Roman Coronagraph) with an inner working angles (IWA) \citep{Traub-Oppenheimer:2010,CarrionGonzalez-etal:2021} designed to demonstrate space coronagraphy at sensitivity levels of Jovian-mass planets and faint debris disks in reflected starlight\footnote{\url{https://www.jpl.nasa.gov/missions/the-nancy-grace-roman-space-telescope}}. In addition, long-baseline optical interferometry may also be used to resolve gravitational microlensing events  \citep{Delplancke:2001,Cassan-etal:2022}. Therefore, the modern generation of instruments may already have the capabilities needed to observe and image the evolving morphology of the Einstein rings. Even though some of these space telescopes were not designed for gravitational lensing studies, they have advantages that come from potentially continuous, high-cadence temporal sampling that is possible from space.

As an example, the Einstein ring around $\alpha$ Cen A has an angular radius of $\sim 81$~mas, which is clearly resolvable by several existing telescopes at optical wavelengths. This analysis suggests that, although the current instruments were not designed to observe Einstein rings around nearby stars, the needed capability exists. In any case, the technology is available and soon may be deployed for dedicated observations of the Einstein rings in our stellar neighborhood.

Furthermore, as imaging capabilities improve, several existing facilities may be used to simultaneously detect microlensing events and trigger imaging observations of  resolved stellar Einstein rings, similarly to what was done for microlensing by the Optical Gravitational Lensing Experiment (OGLE\footnote{\url{http://ogle.astrouw.edu.pl/}} \citep{Udalski:1992}), the Korean Microlensing Telescope Network (KMTNet\footnote{\url{https://kmtnet.kasi.re.kr/kmtnet-eng/}} \citep{Kim-etal:2010}), RoboNet \citep{Tsapras-etal:2009}, The Zwicky Transient Facility\footnote{\url{https://www.ztf.caltech.edu/}}  \citep{Bellm-etal:2018,Rodriguez-eatal:2022} and other astronomical facilities.

\subsection{Estimating the occurrence rate}
\label{sec:occrate}

How often can we see the lensing events that produce Einstein rings with sizes that may be resolvable with available or near future instruments (i.e., see Table~\ref{tb:tel-cap})? Specific predictions \citep{Kervella2016,Mustill-etal:2018,DiStefano:2008,Kluter-etal:2018,Kiroglu-etal:2022} notwithstanding, many such events will be discovered by chance by extensive monitoring of candidate lenses for an extended period of time. Would such monitoring be practical and worthwhile?

There have been prior estimates of the frequency of lensing events \citep{Liebes:1964,Paczynski:1996a}). In particular, \citep{Liebes:1964} estimates that for any given nearby star with a globular cluster of object stars in the background, events involving light amplification in excess of 100 will occur, on average, once every ten thousand years and last about 3 days. If this is the case, then serendipitous discoveries of lensing events will be rare, and a monitoring campaign of even a large number of candidate lenses will likely remain fruitless. The focus, in this case, would have to shift to specific events that can be predicted with reasonable certainty.  Modern astrometric catalogues (i.e, the Gaia Catalogue of Nearby Stars \citep{Smart-etal:2021}) are at the few tens of $\mu$as level accuracy \citep{Lindegren-eta:2021} that is needed to predict likely stellar conjunctions and on-going technology developments may may soon be available for an astrometric search for Earth-like exoplanets \citep{Shao-etal:2018,Ji-etal:2022,Shao-etal:2022}.

We first consider the dynamics of a lensing event, as expressed by (\ref{eq:Amp10+}). We use the relative proper motion between the lens and the source,  $\mu_{\rm rel}$, as seen by the observer, and assume a vanishing impact parameter, $u_0=0$. Under such conditions, the duration of a lensing event, $ \Delta t_{\tt E}$, or the time for the source to cross the entire diameter of the Einstein ring  of the lens is  
{}
\begin{eqnarray}
\Delta t_{\tt E}&=& 
\frac{2\theta_{\tt E}}{\mu_{\rm rel}}=6.59
\Big(\frac{M_{\tt L}}{0.1M_\odot}\Big)^\frac{1}{2}\Big(\frac{\overline z}{10~{\rm pc}}\Big)^\frac{1}{2}\Big(\frac{1~{\rm ''/yr}}{\mu_{\rm rel}}\Big)~{\rm days}.
\label{eq:n-lens-dura}
\end{eqnarray}

Note that this is the duration of the entire lensing event. A full Einstein ring would only be visible if the impact parameter is very small, and even then, only briefly. The actual duration of the appearance of the full Einstein ring depends also on the size of the source, since it corresponds to the observer traversing an image of that source, as projected by the lens into the image plane. For a source with a physical radius $R_{\tt s}$ and diameter of (\ref{eq:rE}), assuming again a ``bull's eye'' event with zero impact parameter, this duration can be estimated as
 {}
\begin{eqnarray}
\Delta \tau_{\tt ER}&=&
\frac{2(\bar z/z_0)R_{\tt s}}{\bar z \mu_{\rm rel}}\equiv 
\frac{2R_{\tt s}}{z_0 \mu_{\rm rel}}=2.93
\Big(\frac{R_{\tt s}}{0.1R_\odot}\Big)
\Big(\frac{10~{\rm kpc}}{z_0}\Big)
\Big(\frac{1~''/{\rm yr}}{\mu_{\rm rel}}\Big)~{\rm sec},
\label{eq:n-lens-dura}
\end{eqnarray}
an event duration that is compatible with current generations of imaging sensors but which should also serve as guidance for the design of future instruments.

Let us now estimate the rate of occurrence of such lensing events.  For that we consdier the case of a lens situated in front of a dense stellar cluster and make some basic assumptions about the event geometry as shown by Figs.~\ref{fig:imaging-geom}--\ref{fig:dir-image}.

As seen by an observer, a lens with a linear width of $2\theta_{\tt E}\bar z$, moving with the relative proper motion of $\mu_{\rm rel}$ with respect to a distant star field, in the time interval of $\Delta t$ subtends an area of $\bar A=2\theta_{\tt E}\bar z^2\mu_{\rm rel} \Delta t$ on the lens plane. To compare this value to the surface density of a distant starfield, we project this area on to the source plane as $A_0\simeq({z_0}/{\overline z})^2\bar A$, which now is the area traversed by the lens' projection on the source plane with a dense starfield.
To calculate the event rate, we need to know the surface density $\sigma_{\tt bg}$ of stars 
in such a background. For that, we 
use the values of a surface mass density of  
very dense stellar clusters of $\sim 10^{5}~M_\odot \, {\rm pc}^{-2}$  \citep{Hopkins-2010,Grudic-etal:2019}. Assuming the cluster consists entirely of Sun-like stars,  this implies a surface number density of background stars of $\sigma_{\tt bg}=10^5({M_{\tt s}}/{M_\odot})~{\rm pc}^{-2}$, that gives an estimate of the number of lensing events in a time interval $\Delta t$:
{}
\begin{eqnarray}
N_{\tt LE}&=&\sigma_{\tt bg}A_0=
\sigma_{\tt bg}2\theta_{\tt E}z_0^2\mu_{\rm rel} \Delta t\simeq 
4.24 \Big(\frac{\sigma_{\tt bg}}{10^5~{\rm pc}^{-2}}\Big)
\Big(\frac{M_\odot}{M_{\tt s}}\Big)
\Big(\frac{M_{\tt L}}{0.1 M_\odot}\Big)^\frac{1}{2}
\Big(\frac{z_0}{10~{\rm kpc}}\Big)^2\Big(\frac{10~{\rm pc}}{\overline z}\Big)^\frac{1}{2}
\Big(\frac{ \mu_{\rm rel}}{1~''/{\rm yr}}\Big)\Big(\frac{\Delta t}{\rm yr}\Big).~~~
\label{eq:n-lens}
\end{eqnarray}

This analysis confirms that the frequency of such lensing events warrants further effort. Regular monitoring, e.g., using a wide-field instrument with a coronagraph such as that of the upcoming Nancy Roman space telescope (see Table~\ref{tb:tel-cap}), of a modest number of foreground stars with dense backgrounds may yield several such lensing events annually, which could then be studied in detail, including the evolving morphology of the emerging Einstein ring, through follow-up observations with higher resolution (both astrometric and temporal) instruments, especially those that are being developed for exoplanet studies \citep{Carrion-Gonzalez-eat:2021}.

Of course actual occurrence rates must be calculated for specific lensing candidates given the background in front of which they appear \citep{DiStefano:2008,Kiroglu-etal:2022}, their proper motion across the sky \citep{Bovy:2009}, and the density and nature of the background (i.e., dense star clusters, white dwarfs, black holes). In addition, the specific motion of an observing facility or telescope network has to be accounted for \citep{Zhu-Gould:2016,Calchi-Novati:2016,Street-Bachelet:2019,Carrion-Gonzalez-eat:2021}.

\subsection{Modeling tools}
\label{sec:m-tools}

The results presented in the previous subsections and in the referenced papers are practical and actionable in the sense that they can be implemented as numerical computer codes, to simulate the behavior of gravitational lenses and model imaging \citep{Turyshev-Toth:2021-quartic,Turyshev-Toth:2021-all-regions,Turyshev-Toth:2021-STF-moments,Turyshev-Toth:2022-broad_SNR,SGL2021g,Toth-Turyshev:2020,SGL2021h}.

We have developed a growing set of software tools, intended to be shared with the wider community once they reach sufficient maturity. These tools, mostly {\tt C++} language programs, implement the PSF of a monopole gravitational lens and a monopole lens perturbed by multipole moments. We implemented several algorithms for the PSF.

In Sec.~\ref{sec:im-form-proc} we discussed imaging regions, specifically the strong and weak interference regions. We indicated that the weak interference region approximation is required when the displacement of the major and minor images from the Einstein ring becomes significant. When viewed through a diffraction-limited telescope, this displacement can be ignored when $\rho_0\lesssim {\overline z}(\lambda/2d)$, but becomes noticeable when $\rho_0$ is larger.

Closer to the optical axis, the major and minor images of a distant point source widen into arcs. This is not captured by the weak interference region approximation. When displacement from the optical axis remains much larger than the spatial wavelength of the PSF, $\rho_0\gg \lambda/\theta_{\tt E}$, the weak interference region approximation remains valid; for smaller values of $\rho_0$ the strong interference region approximation must be used.

Conversely, at very large distances from the optical axis, light amplification becomes insignificant, and eventually, when $\rho_0 > 2r_g/(\lambda/2d)$, the displacement of the image due to the gravitational bending of light can no longer be observed within the telescope's diffraction limit. These considerations can be readily adapted for extended sources by integrating over the surface of the source.

\begin{figure}[t!]
\begin{center}
\includegraphics{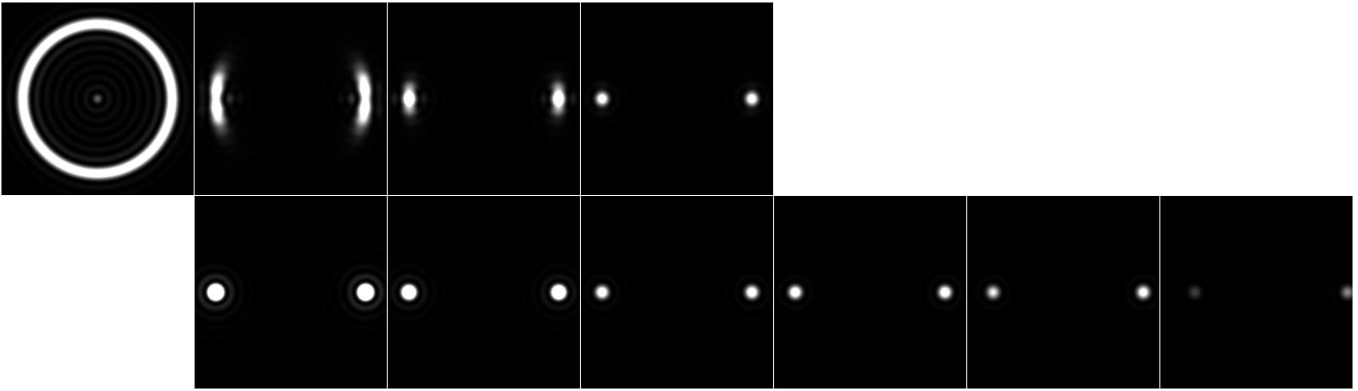}
\end{center}
\caption{\label{fig:tests} Morphology of the evolving Einstein ring around a $1~M_\odot$ lens formed by the light from a point source and observed at $\lambda=1~\mu$m
by a telescope at 650~AU from the lens. Results for the strong and weak interference region approximations are shown by top and bottom rows, correspondingly. On the optical axis (1st column) the weak approximation is divergent. At $\rho_0=1$~m and $\rho_0=2$~m (2nd, 3d columns) prominent arcs are visible but these are only modeled by the strong approximation; the weak approximation shows point-like images. At $\rho_0=5$~m (4th column), the two approximations become indistinguishable. At much greater distances, the strong approximation becomes computationally costly and must be abandoned, but even at $\rho_0=10^6$~m (5th column) it would yield indistinguishable results. By $\rho_0=10^8$~m, the images begin to drift from the nominal Einstein ring position (6th column). Finally, at $\rho_0=3\times 10^8$~m we also begin to see that the minor image is noticeably fainter. Note that the brightness of images at greater $\rho_0$ has been enhanced for visibility on the printed page or viewing screen.
}
\end{figure}

In imaging scenarios that involve either the solar gravitational lens or stellar gravitational lenses in our Galactic neighborhood, at optical or near-IR wavelengths, the strong interference region an weak interference region approximations overlap and yield identical results across several orders of magnitude in the value of the $\rho_0$ parameter. This is convenient, indeed critical for numerical efficiency, as the computational cost of the strong interference approximation rapidly increases (requiring ever finer integration steps) with increasing distance from the optical axis. Thankfully, very close to the optical axis, at $\rho_0\gg \lambda/\theta_{\tt E} +r_{\tt s}$, the weak interference approximation becomes valid, its results in fact indistinguishable from those obtained using the approximation for the strong interference region. This is demonstrated for a point source in Fig.~\ref{fig:tests}. Using the solar gravitational lens as an example, this figure shows that when $\rho_0\gtrsim 40\lambda/\theta_{\tt E}$, the two formalisms yield identical results. For the solar gravitational lens at ${\overline z}=650~$AU and $\lambda=1~\mu$m, this means a mere 5 meters from the optical axis! Being able to rapidly switch to the weak interference region approximation makes it possible to model even the lensing of large sources.

The PSF can be convolved with source images using direct or Fourier convolution to model the formation of images projected by a lens. It can be used to model image reconstruction (deconvolution) and estimate the signal-to-noise ratio and the requisite integration times in the presence of realistic noise. The gravitational lens PSF can also be convolved with that of a thin-lens telescope to simulate images that a telescope sees when looking at a distant object through the lens. Lastly, using an adaptive integration scheme, we can apply this combined PSF to model light from extended sources, such as a stellar disk.

With these tools at our disposal, we can calculate wavelength-dependent models of lensing events. Though the tools were developed primarily for use with the solar gravitational lens, they can also be used with stellar lenses with only a minor penalty in numerical stability despite the vastly different geometry. The tools, though still in development, are sufficiently efficient to produce not just still image reconstructions but also animations of lensing events, including lensing of extended sources.

\subsection{Modeling a lensing event}
\label{sec:mod-lensevent}

\begin{figure}[t!]
\begin{center}
\includegraphics{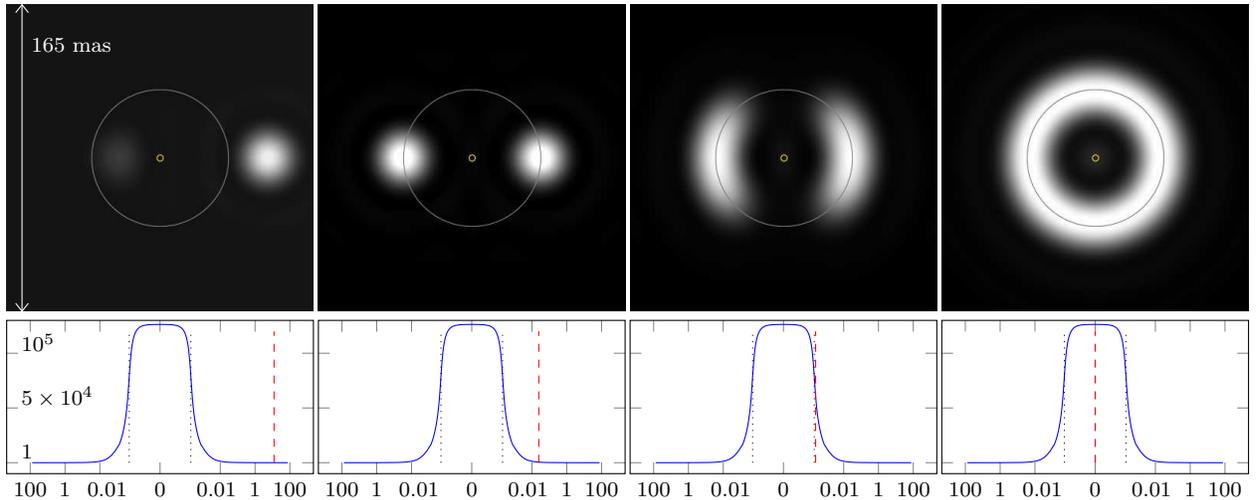}
\end{center}
\caption{\label{fig:SGL-im-sensor}
Evolution of the Einstein ring, viewed through a diffraction-limited telescope with $d=10$~m at the wavelength $\lambda=1~\mu$m. The $M_\odot$ lens is at $\overline z=6$~pc from the telescope, moving at 30~km/s proper velocity, amplifying light from a Sun-like star at $z_0\sim 8$~kpc. The four images shown correspond to angular separations between source and lens at $\sim$35~mas (minimal light amplification), $\sim$70~$\mu$as, $0.7~\mu$as and conjunction. Thin circles show the outline of the lens and its Einstein ring. Plots show light amplification, with the horizontal axis representing, on a logarithmic scale, the number of days before and after the peak event. Dashed vertical lines in the plots show the position of the source with respect to the lens. Dotted vertical (blue) line marks the boundaries of the geometric projection of the lensed object, i.e., the $\sim$35~sec interval during which a full Einstein ring becomes visible. For an animation, see \protect\url{https://sglf.space/sterings.php}.
}
\end{figure}

The process of the Einstein ring formation around such a lens is well understood, including the impact of the spherical aberration on the its imaging properties \citep{Toth-Turyshev:2020}. Although the thickness of the Einstein ring is not going to be resolved with a typical telescope, the ring's angular extent and the evolving lengths of the individual arcs will be available as well as their contours trimmed by the diffraction-limited telescope performance (see Fig.~\ref{fig:SGL-im-sensor}), thus providing us with a set of important observables.

Thinking of the gravitational lens as an imperfect projector, it projects a blurry image of the distant source onto the plane where the observing telescope is situated. We can describe the evolving morphology of the view seen by the telescope in terms of its distance from this blurred image in the image plane. The image is centered on the primary optical axis: the line that connects the center of the source with the center of the lens. When the telescope is at a very large separation from the optical axis, $\rho_0\gg r_{\tt s}$, expression (\ref{eq:FI-go-Int}) characterizes the unamplified light received from a distant target. Approaching the optical axis but still keeping $\rho_0> r_{\tt s}$, we enter the weak interference region of the lens, with the intensity given by (\ref{eq:FI-ir+-Int}).

When the telescope approaches the projected image, $\rho_0 \sim  r_{\tt s}$, we enter the strong interference region where the intensity of light at first is described by (\ref{eq:P-blur*off4}).  Finally, as the telescope reaches the vicinity of the optical axis,  $0\leq \rho_0 \leq  r_{\tt s}$, the intensity now is given by (\ref{eq:intense2*}). These expressions allow us to study the process of  the Einstein ring's formation as the function of distance with the primary optical axis, $\rho_0(t)$, which parameterizes the telescope's alignment with respect to the primary optical axis. Their appearance notwithstanding, these expressions can be integrated numerically and form the basis of our ability to create simulations of the view seen by an imaging telescope that is looking at a distant object through a stellar gravitational lens.

Having developed the needed modeling tools, we are ready to model actual lensing events as seen through a realistic astronomical telescope: an observation that combined conventional microlensing with imaging. We can then focus our attention on what we can learn from such an encounter. Applying what we have, we can look at actual stars in our solar system's neighborhood, evaluating their characteristics in light of available instruments. We can also establish realistic estimates on the statistical frequency of such lensing events.

Using our simulation tools, we modeled a hypothetical lensing event that involves a Sun-like star situated at 6~pc from the solar system, coming into conjunction with a background star, also Sun-like, at a distance of 8~kpc. We assumed that the proper motion of the lens is 30 km/s, a value we picked because it happens to be the Earth's orbital velocity around the Sun, and as such, a typical value for the proper motion of a nearby star as seen from an observatory on Earth, in geocentric or in heliocentric orbit.

This encounter is depicted in Fig.~\ref{fig:SGL-im-sensor}. We present a simulation of the diffraction-limited view by a telescope with a 10~m aperture, operating at the wavelength $\lambda=1~\mu$m, along with a light amplification curve that is familiar from microlensing models and observations.

The entire duration of the depicted event is 150 days, but for much of that time, there is negligible light amplification. The first panel in Fig.~\ref{fig:SGL-im-sensor} shows the event approximately 12 days before maximum brightness: the primary image of the source (appearing on the right) remains unamplified, but a faint secondary image emerges to the left of the lens. Light amplification increases gradually, reaching $\sim$100 approximately 3 hours before conjunction.

The second panel in Fig.~\ref{fig:SGL-im-sensor} shows the event 35 min before conjunction, at which point light amplification exceeds $\sim$500.
At this point, the amplification increases very rapidly, as the telescope approaches the region in the image plane where the projected image of the source is located, $\rho_0 \to r_{\tt s}$.

 The third panel in Fig.~\ref{fig:SGL-im-sensor} shows the view by this telescope shortly before it is about to enter this region, approximately 20 sec before conjunction. At this point, light amplification is nearly $6\times 10^4$.
Under these conditions, the full Einstein ring forms a few seconds later and remains visible (albeit with varying intensity) for approximately 30 sec. At conjunction, light amplification is in excess of $1.25\times 10^5$
(shown in the last panel of Fig.~\ref{fig:SGL-im-sensor}).

\begin{figure}[t!]
\begin{center}
\includegraphics{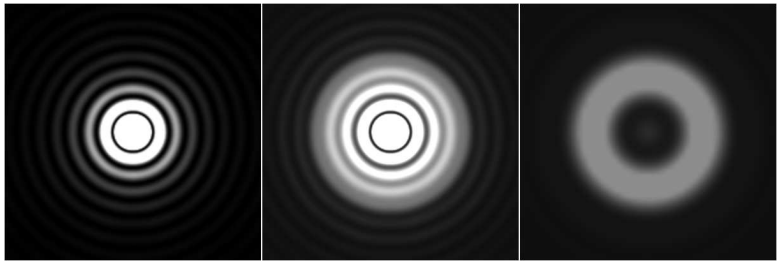}
\end{center}
\caption{\label{fig:lensing}Simulated view of a lensing event with a lens at $\sim$6~pc, lined up with a background star at $\sim$8~kpc. Left: the foreground star as viewed through a thin lens telescope with a 10~m aperture at $\lambda=1~\mu$m. Center: the same foreground star, with the Einstein ring of a background star (as seen in the last image in Fig.~\ref{fig:SGL-im-sensor}). Although the Einstein ring is fainter and the Airy pattern associated with the foreground star obscures it, the ring is nonetheless clearly visible. Right: Just the Einstein ring. Apodization or a stellar coronagraph or starshade may make it easier to distinguish the Einstein ring from the lens.}
\end{figure}

These simulated images were obtained by numerically integrating the PSF of the gravitational lens, convolved with the PSF of the imaging (thin lens) telescope. It is notable, however, that the light amplification curve that we obtained follows very closely the values predicted by the closed-form expression (\ref{eq:power0}), formulated in terms of the well-known elliptic integral. Eq.~(\ref{eq:power0}) is therefore confirmed as a powerful and accurate predictor of the light amplification curve of a monopole gravitational lens acting on a uniformly illuminated disk as the source.

The images in Fig.~\ref{fig:SGL-im-sensor} do not depict the lensing star. In realistic imaging scenarios, we must consider light contamination from the lens. When it is not blocked by a coronagraph, the result may appear to that shown in Fig.~\ref{fig:lensing}, with the Einstein-ring overlapping the concentric rings of the Airy disk that correspond to the lens. Apodization may reduce the prominence of the Airy pattern, but at the cost of reduced resolution that may make it harder to distinguish the Einstein ring from the lens. Similarly, a telescope equipped with an internal coronagraph and a Lyot stop may successfully block light from the lens, but at the cost of reducing the telescope's resolution.

The optimal combination of optical properties, observing wavelength, and the use of coronagraph may depend on the specific observational targets and anticipated lensing events. The choice or design of instruments used for such observations may be based on considerations that include the range of Einstein ring angular radii, lens star luminosities and lens star spectra.

\section{The case of $\alpha$ Centauri A in 2028}
\label{sec:acent}

In the preceding section, we estimated the statistical rate of occurrence of lensing events involving nearby stars, which may result in a resolvable Einstein ring, allowing us to study the temporal evolution and geometry of the event, and obtain high resolution spectra of the lensed object. We treated the events as unpredictable, estimating their statistical rates of occurrence. However, there are cases when conjunctions can, and have been, predicted with significant probabilities \citep{Kervella2016,Mustill-etal:2018}, including one in 2028. Uncertainties in the relevant astrometric observables imply that we do not know for certain if this encounter will take place, but there is a chance that $\alpha$ Cen A will be in conjunction with a distant red supergiant in 2028 and an observable Einstein ring may appear.

\subsection{The predicted conjunction in 2028}

The K-magnitude 7.8 star {\tt 2MASS 14392160-6049528}, a presumed red supergiant located at $\approx 10^4$ light years from Earth, is expected to come within $0.015\pm 0.135''$ of $\alpha$ Cen A in May 2028. This stellar conjunction may offer an opportunity to observe a predictable Einstein ring occurrence.

Little is known \citep{vizier} about the distant star\footnote{See \url{http://vizier.u-strasbg.fr/viz-bin/VizieR-5?-out.add=.&-source=2246&-c=219.839915\%20-60.831290,rs=0.5}
} that will be the lensed. As it is described as a red supergiant, we endeavored to model {\tt 2MASS 14392160-6049528} as a star with a stellar radius of $\sim 0.72$~AU, located at $10^4$~light years. We note that due to its enormous size, despite its great distance such a large star subtends an angle of nearly half a mas. Therefore, even a ``near miss'' conjunction between $\alpha$ Cen A and {\tt 2MASS 14392160-6049528} may result in a partial or full Einstein ring becoming visible.

On the other hand, light amplification is less dramatic when the object being lensed is large. The FOV of a gravitational lens is very narrow. At any given moment in time, the lens will show a small portion of the lensed object. Although light from this visible cross-section will be greatly amplified, compared to the light of received form the entirety of the object (including parts that fall outside the FOV of the gravitational lens), light amplification will be proportionately less. Yet given the intrinsic brightness of {\tt 2MASS 14392160-6049528}, we can still expect a significant event, with an Einstein ring that may become substantial even against the contrast of the bright light of $\alpha$ Cen A, a nearby Sun-like star.

\subsection{Modeling the 2028 encounter}

Adapting our existing gravitational lens model, we assumed that {\tt 2MASS 14392160-6049528} is located at $10^4$ light years from us, and its light is lensed by a Sun-like star at the distance of $\alpha$ Cen A. (The difference between the masses of $\alpha$ Cen A and the Sun is not significant compared to the other approximations in this estimation.)

\begin{figure}[t!]
\begin{center}
\includegraphics{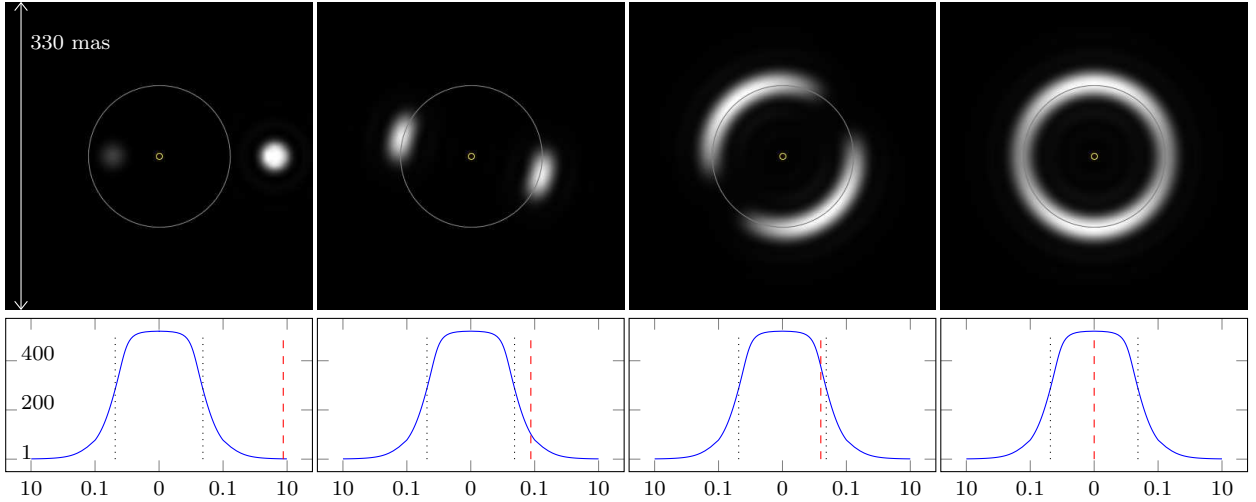}
\end{center}
\caption{\label{fig:ACENTS5D}
Evolution of the Einstein ring during a possible that might take place between $\alpha$ Cen A as the lens and the red supergiant {\tt 2MASS 14392160-6049528} in May 2028. Depicted is a glancing encounter with an angular impact parameter of 0.2~mas. Plots show light amplification, with the logarithmic horizontal axis measuring the number of days before and after the peak event. Angular distance from closest approach 75 mas, 0.75 mas, 0.15 mas and 0 mas, respectively. Dashed vertical lines show the position of the source with respect to the lens. Dotted vertical (blue) lines mark the boundaries of the geometric projection of the lensed object, i.e., the interval in which a full Einstein ring becomes visible. For an animation, please see \protect\url{https://sglf.space/acents.php}.
}
\end{figure}

Rather than modeling a perfect, ``bull's eye'' encounter (that is, an angular separation of 0 at peak light amplification) we assumed an offset, or glancing, encounter, with a minimum angular separation of 0.2~mas between $\alpha$ Cen A and {\tt 2MASS 14392160-6049528}. Given the enormous size of {\tt 2MASS 14392160-6049528}, this still puts us well inside its projected image in the image plane, even though the modeled observing telescope never reaches the vicinity of the primary optical axis.
Such an offset, glancing encounter offers an interesting case study as we look at the temporal evolution of the Einstein ring. Not only do arclets emerge, but they also rotate, following the relative direction of {\tt 2MASS 14392160-6049528} with respect to $\alpha$ Cen A. (Note that a similar effect of rotating gravitational arcs around a microlens, Gaia19bld9, made with the PIONIER\footnote{
Precision Integrated-Optics Near-infrared Imaging ExpeRiment (PIONIER),\\\url{https://www.eso.org/public/teles-instr/paranal-observatory/vlt/vlt-instr/pionier/}} instrument at the VLTI  was discussed in \citep{Cassan-etal:2022}.) This is shown in Fig.~\ref{fig:ACENTS5D}. As the Earth's orbital motion dominates the anticipated alignment, the time span between the first and last panel in this figure corresponds to $\sim 6$ days assuming that the observing telescope moves across the FOV at the orbital speed of the Earth, $\sim$30~km/s. The next-to-last image (two arcs almost joined to form a complete Einstein ring) would occur $\sim$15 minutes before peak light amplification (the final panel). Such a rotation of gravitational arcs provides information on the lens-source relative motion and could be used to improve the proper motion of the source.

We also evaluated the light amplification that is expected during this event. As expected, due to the very large size of {\tt 2MASS 14392160-6049528}, light amplification is more modest, peaking at $\sim 520$. (A perfect, ``bull's eye'' encounter would yield a light amplification of $\sim 670$.) Nonetheless, this is significant. The K-magnitude of the Einstein ring increases from the value of 7.8 (that of the unamplified star) to $\sim 1$, which makes the ring comparable in brightness to $\alpha$ Cen A itself. Another way to estimate the brightness of the Einstein ring is by noting that even if {\tt 2MASS 14392160-6049528} is a fairly cool red giant with a surface temperature of 3500~K, given a radius of 0.72~AU and a distance of $10^4$ light years, from (\ref{eq:Pexo-pdsl0}) we get an unamplified power density of $B_0\sim 0.011\times 10^{-9}$~W/m$^2$; amplified by a factor of 520, using (\ref{eq:Pexo-pdsl}) we obtain $B_{\tt ER} = 5.77\times 10^{-9}$~W/m$^2$ for the resulting Einstein ring, which amounts to more than 20\% of the brightness of $\alpha$ Cen A (see Table~\ref{tb:pdreceive2}). Given the brightness of the ring and its large angular separation from the lens star, it represents a readily available, prime candidate for a stellar Einstein ring observation.

\subsection{Planetary detection}

Microlensing events have long been recognized as tools for discovering exoplanets in lensing systems. Too faint to be seen on their own, a fortuitous conjunction between a planet and a background object, with the planet acting as a lens, can produce a transient event that can help identify the planet's location, measure its mass, and determine its orbit. Indeed, there have been attempts to predict possible planetary lensing events of this nature \citep{Mustill-etal:2018,Beichman-etal:2020}.

While a planetary Einstein ring may be too small to be resolvable by existing or planned astronomical instruments, observing a planetary lensing event in the context of its host system can reveal additional information. Depending on the lensing system geometry, the planetary event may happen either before or after the ``main'' event, when either the primary or the secondary image of the distant object is further amplified by the planetary gravitational lens.

\begin{figure}[t!]
\begin{center}
\includegraphics{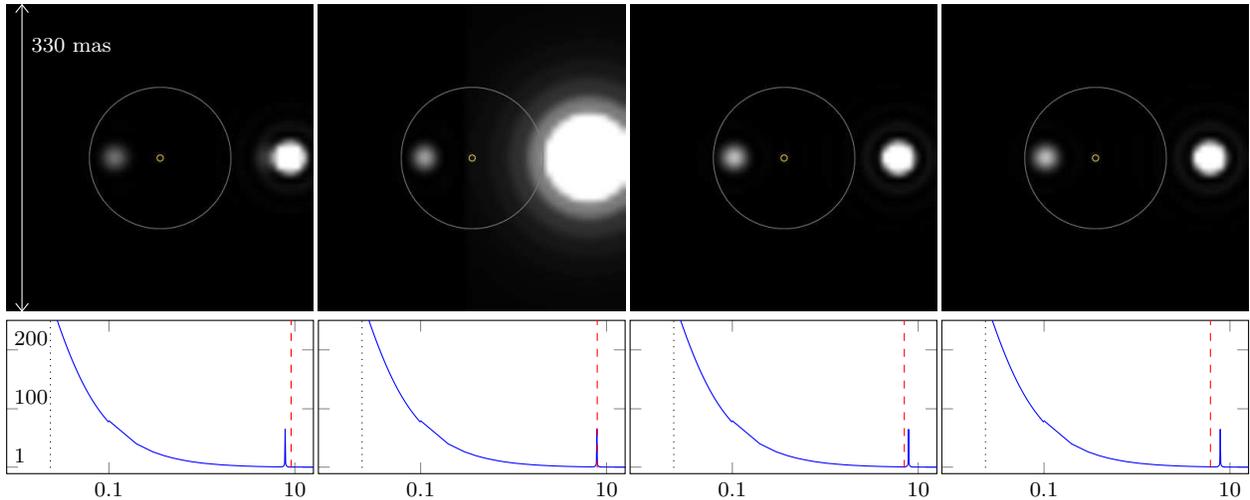}
\end{center}
\caption{\label{fig:jevent}Modeling a hypothetical planetary lensing event approx. 10 days before the ``main'' event of the $\alpha$ Cen A encounter with {\tt 2MASS 14392160-6049528} in 2028. Horizontal axis is the time in days. These images correspond to the first panel in Fig.~\ref{fig:ACENTS5D} with the addition of the modeled planet and its contribution to light amplification of the primary lensed image. Although the planetary Einstein ring is not resolvable, its position relative to the host star constrains the planet's orbit.}
\end{figure}

We endeavored to model a hypothetical planetary lensing event in the context of the anticipated 2028 $\alpha$ Cen A conjunction. To do so, we assumed that before the event, as it was modeled in the preceding subsection, there a ``perfect'' conjunction involving a planet orbiting $\alpha$ Cen A. Such a conjunction yields a clearly identifiable secondary peak in the light amplification curve of the lensing system (Fig.~\ref{fig:jevent}). The peak light amplification of this event is $\sim 75$, and light amplification in excess of 10\% of this peak may be present for up to 3.5 hours.

Planets with masses as small as $M_\oplus$ may be detectable through microlensing. As light amplification of the lens is approximately scales as $\propto \sqrt{M_{\tt L}}$, lenses much smaller than the modeled lens\footnote{
%We should stress
Note that for illustrative purposes and also because our computational tools, still in development, show greater numerical stability for massive lenses, we chose a massive planet ($10M_{\tt J}$). However, even planets
smaller than $M_\oplus$ will be detectable during this lensing event.}, including lens masses $\lesssim M_\oplus$, will be easily visible if they come in conjunction with the lensed star. Furthermore, the $\alpha$ Cen A 2028 encounter is far from an ideal case: the large size of the lensed object offers reduced light amplification. When lensing involves a more compact light source, light amplification is more prominent and even smaller planets may be more easily detectable.

It is, of course, well known that planetary lensing produces such photometric spikes in the light amplification curves during lensing events. As seen in Fig.~\ref{fig:jevent}, resolved imaging adds significant new information. The position of the image during brightening event can inform on the position of the lensing planet within its host solar system. The magnitude of the brightening can be assigned unambiguously to either the primary or the secondary image of the lensed star with respect to the lensing host star, which can lead to tighter, more reliable estimates on the mass of the planet responsible for this spike help constrain its orbital parameters (i.e., semi-major axis, inclination, etc.).

Finally, we should mention that even if planetary lensing events are not specifically predicted in advance, they can be anticipated to occur before or after stellar lensing events. Therefore, it is beneficial to monitor potential candidate solar systems not just for Einstein rings appearing around the host star, but also for spikes of light amplification before or after a host star lensing event. Depending on the planet's semi-major axis and its position along its orbit, such planetary events may take place as much as several weeks before or after peak lensing by the star. Detection of such events and pinpointing their location in the sky relative to the lensing star can lead to either confirmation of the orbital parameters and masses of previously known exoplanets or new exoplanet discoveries around nearby stars.

\subsection{Photometric imaging with a network of telescopes}

The case of the 2028 $\alpha$ Cen A conjunction with {\tt 2MASS 14392160-6049528} closely resembles the scenario we depicted in Fig.~\ref{fig:dir-image}, which shows a projected image of the source that is much larger than the Earth. An $R_{\tt s}=0.72$~AU red giant at $z_0=10^4$~ly, projected by a lens at ${\overline z}=4.37$~ly, results in an image that is $\overline z/z_0 2R_{\tt s}\simeq9.41 \times 10^4$~km wide, which is more than 7 times the Earth's diameter. During the lensing event, a potentially large number of observing facilities  worldwide ($\sim 40$ or more), augmented by observatories operating in space, will traverse this image, presenting an interesting imaging opportunity.

While a large observatory may see the evolution and morphology of the Einstein ring, smaller observatories specialized in microlensing can still make photometric observations of the event (as discussed in \citep{Zhu-Gould:2016}). Every such observation amounts to sampling the projected image of {\tt 2MASS 14392160-6049528} at a specific location. If the image is approximately static during the lensing event (as we would expect it to be, considering the very large size of the source), a series of photometric observations by a single telescope amount to sampling the image along a one-dimensional track.

If a network of telescopes is used, on the other hand, each of these telescopes traces an approximately parallel track across the image of {\tt 2MASS 14392160-6049528}. The result can be a series of samples, or ``pixels'', that together constitute a resolved image of a band traversing the visible disk of {\tt 2MASS 14392160-6049528}. Adding
several space-based observatories may increase the  separation between the pixels, potentially covering the entire image and more. The modeling and analysis tools to recover resolved images from similar observations were developed by us for a resolved exoplanet imaging with the solar gravitational lens \citep{Turyshev-Toth:2020-extend,Toth-Turyshev:2020,Turyshev-Toth:2022-broad_SNR} and are available for such efforts.

Given the large size of the projected image of {\tt 2MASS 14392160-6049528} and the modest speed ($\sim$30~km/s) at which the Earth traverses this image, the event may last up to half an hour or longer, depending on its precise geometry. The frequency at which an observing telescope can make photometric measurements may vary wildly between instruments but it is reasonable to assume that tens, perhaps even hundreds of measurements can be performed by a single instrument during such a period of time. A network of instruments, therefore, may obtain a kilopixel or higher resolution image of the surface of {\tt 2MASS 14392160-6049528}, precisely measuring its diameter and variability of surface brightness. This represents a unique opportunity to study the physics of a red supergiant.

\begin{figure}
\begin{center}
\includegraphics[scale=0.5]{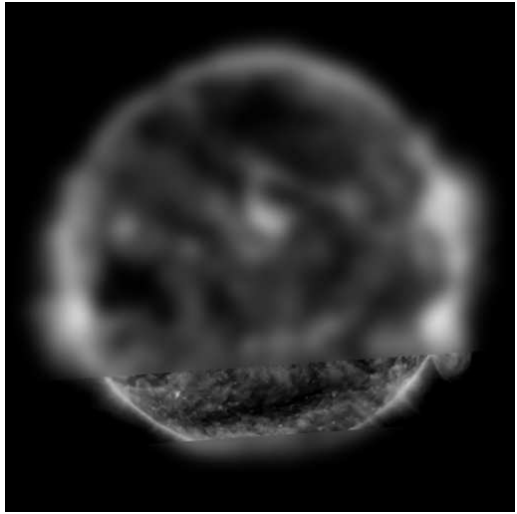}
\end{center}
\caption{\label{fig:sunband}Simulation of an image of {\tt 2MASS 14392160-6049528} that might be recoverable using a combination of a network of terrestrial and space-based telescopes. The projected image of the star is much larger than the diameter of the Earth. As the Earth traverses this image, terrestrial telescopes can obtain a high resolution sample along the Earth's path. Outside the Earth, using a small handful of space-based assets, the rest of the star may be reconstructed at a much lower resolution.
}
\end{figure}

As stellar gravitational lenses are not perfect, especially considering their spherical aberration \citep{Turyshev-Toth:2020-extend}, it is natural to wonder if any image obtained this way would be subject to substantial blur. As we discussed in \citep{Turyshev-Toth:2022-broad_SNR}, when the sampling locations are widely spaced, the blur due to neighboring pixels becomes negligible, and it is not necessary to use deconvolution \citep{Toth-Turyshev:2020} to recover a sharper image. This is fortunate also because when a diverse network of telescopes is used for coordinated observations through the atmosphere, one can naturally expect variabilities in observed brightness that are difficult to model. The related measurement uncertainties contribute to noise, which would be  amplified by deconvolution \citep{Turyshev-Toth:2020-extend,Toth-Turyshev:2020}. However, as the sampling separation will be quite large, no such issues will arise \citep{Turyshev-Toth:2022-broad_SNR} and  good quality images of the source may be obtained.

This is demonstrated in Figure~\ref{fig:sunband}. To obtain this image, we convolved a randomly chosen image of our own Sun, properly scaled and used a stand-in for {\tt 2MASS 14392160-6049528}, with the PSF of the $\alpha$ Cen A gravitational lens. The convolution yielded no appreciable degradation in image quality, as predicted. In comparison with our earlier models of exoplanet imaging using the solar gravitational lens, the reason is evident: the projected image of {\tt 2MASS 14392160-6049528} is several ten thousand times larger than the projected exoplanet images we considered. Even at megapixel resolution, sampling locations of would be up to 100~km apart, much too far for any appreciable blur to appear between neighboring pixels.

We then used standard desktop image processing tools to show a high resolution path across this image (representing a network of terrestrial telescopes) with the rest of the star shown at substantial blur, corresponding to sparse sampling by a small number of space-based telescopes. It is evident from the simulated image that the basic geometric properties of the distant star are recoverable, along with large-scale features such as limb darkening, which can provide information about the stellar atmosphere. The high-resolution path across the star's surface can reveal details that characterize stellar activity, providing perhaps an unprecedented opportunity to study the physics of this red giant.

\section{Discussion and Conclusions}
\label{sec:disc}

Every star is destined to have a ring: an Einstein ring, that is. Strong gravitational lensing by stars in our stellar neighborhood offers a unique method to study distant objects and regions of interest. As given by (\ref{eq:alpha-mu}), the position of the Einstein ring around a star is determined by the distance to the star and its mass. Such lenses offer a unique opportunity to study the light source and the background at resolution that is significantly higher than what is currently available  with modern telescopes \citep{Mao:2012}. To take full advantage of this capability offered by stellar gravitational lenses, it is necessary to resolve the Einstein ring from the lensing star, as opposed to merely observing a change in brightness of an unresolved source.

Lensing events involving nearby stars have already been predicted using existing data in astrometric catalogues. As such catalogues are refined in terms of the astrometric precision, number of stars, and smaller stellar magnitudes and if done by appropriately-trained computer algorithms, we expect more such predictions in the near future. On the other hand, it is also possible to predict the statistical rate of occurrence of unforeseen lensing events. Our preliminary analysis suggests that for foreground stars acting as lenses with dense clusters in the background, lensing events that produce observable Einstein rings may occur at a rate of one such event every year. Therefore, monitoring a number candidate lenses can reliably yield lensing events at a rate of several events a year.

We note that typical Einstein rings for stars in our vicinity,  $\overline z\sim$ 100 pc, would have angular diameters of several ten mas. This is comparable to the anticipated angular size of many exoplanet orbits. Existing or planned coronagraphic instruments designed for exoplanet detection \citep{Carrion-Gonzalez-eat:2021} can, therefore, be easily adapted to observe Einstein rings against the glare of the lens. Einstein rings produced from light of distant stars will be considerably brighter than faint exoplanets, making them easier to observe. Often, the Einstein ring may be as bright as, or brighter, than the lens, so observations even without coronagraphs are also possible. Furthermore, the brightness of these rings implies short integration times, making it possible to study the temporal evolution of the ring.

Lensing events by nearby stars require very close alignment between lens and source. Although difficult, in many cases such events can be predicted using modern astrometric catalogues (i.e., delivered by the ESA's Gaia mission\footnote{\url{https://en.wikipedia.org/wiki/Gaia_(spacecraft)}}, see e.g.,  \citep{Lindegren-eta:2021}) and artificial intelligence algorithms \citep{Smart-etal:2021}. Such predictions have been published \citep{Kervella2016,Mustill-etal:2018}, including the event that is expected occur in 2028, when $\alpha$ Cen A may lens a distant supergiant, {\tt 2MASS 14392160-6049528}. We modeled this event as an example, simulating the formation and evolving morphology of its Einstein ring during a hypothetical, imperfect, ``glancing'' encounter. This and other similar events that are predicted to occur in the foreseeable future with reasonable probabilities should be identified and explored, perhaps in the context of a dedicated space mission.

When stars that can serve as potential lenses are located in front of a background that is a dense stellar neighborhood, even unpredicted lensing events may occur with sufficient frequency to make it worthwhile to monitor them for such occurrences. A campaign looking for Einstein rings may involve survey-type monitoring of candidate lensing stars in search for serendipitous events. It may also be aimed at observing events that have been predicted from astrometric data. Imaging observations can be triggered based on realtime analysis of the photometric light curves monitored by large astronomical facilities or networks of telescopes.

Real-time modeling \citep{Cassan-Ranc:2016} shows that a reasonable prediction of the date and brightness of the peak magnification can be obtained 2-3  days in advance of the peak, providing 48 hours to issue an alert. Furthermore, new generations of microlensing alert telescopes in operation yielded the rate of microlensing event detections of $\sim$  2000 per year \citep{Mao:2012}.
Either way, once a specific lensing target is selected, observations can benefit from the use of a coronagraph, to block light from the lens and improve the ability to detect and analyze the Einstein ring.

Studying the temporal evolution and morphology of an Einstein ring can reveal information about both the lens and the source. For example, the angular diameter of the lens amounts to a direct measurement of the lens mass. If the lensing event includes secondary peaks due to the presence of exoplanets in the lensing system, the precise sky position of the secondary event relative to the lens star can constrain the exoplanet's orbit. The brightening of the (major or minor) image of the distant source can be directly measured, offering a better quality estimate of the lensing planet's mass. Concerning the source, observing the Einstein ring as distinct from the lens allows for spectral analysis of the source without significant light contamination by the lens. Minute changes in the brightness of the Einstein ring around its circumference can help refine astrometric estimates on the relative sky positions of lens and source at levels of accuracy beyond standard astrometric methods.

We also discussed the possibility that as a set of the ground-based telescopes and several space-based facilities moves across the image of the source that is projected by the lens, this network of observing facilities can be used to obtain a resolved image of the source. This can be of substantial importance when the source is of astrophysical interest, as it offers a view of the surface of the source at a resolution that is not otherwise achievable.

Past, present, and proposed missions can image stars, search for exoplanets, study microlensing events. A mission that is dedicated to observe stellar Einstein rings, in contrast, requires instruments capable to actually image these rings, unambiguously determine their brightness, spectra and temporal evolution, and use the resulting information to study the nature, dynamics, and composition of both foreground and background objects. Such a mission may also help establish the orbital geometry of the lensing system if it detects planetary lensing events that are resolved as distinct from the host star. Additionally, the mission would help establish constraints, in the form of quantifying both the geometry and brightness of likely Einstein rings, that may appear and contaminate exoplanet observations.

The results presented here offer motivation to study, describe, and quantify the nature, duration, frequency of occurrence and temporal evolution of stellar Einstein rings. In addition to the possibility of utilizing existing instruments, it should also serve as motivation to study preliminary requirements for a dedicated mission, in an optimally chosen geocentric or heliocentric orbit, to obtain measurements of sufficient quality to be of use for the scientific community. It is time to study, describe, and quantify the nature, duration, frequency of occurrence and temporal evolution of stellar Einstein rings, and establish well-researched, reliable constraints for such an exciting imaging mission.

\section*{Acknowledgments}
This work in part was performed at the Jet Propulsion Laboratory, California Institute of Technology, under a contract with the National Aeronautics and Space Administration. The authors thank Michael Shao of JPL and L. Viktor Toth for helpful discussions.  VTT acknowledges the generous support of Plamen Vasilev and other Patreon patrons.

%\bibliography{SGL-diffract}

\begin{thebibliography}{}
\expandafter\ifx\csname natexlab\endcsname\relax\def\natexlab#1{#1}\fi
\providecommand{\url}[1]{\href{#1}{#1}}
\providecommand{\dodoi}[1]{doi:~\href{http://doi.org/#1}{\nolinkurl{#1}}}
\providecommand{\doeprint}[1]{\href{http://ascl.net/#1}{\nolinkurl{http://ascl.net/#1}}}
\providecommand{\doarXiv}[1]{\href{https://arxiv.org/abs/#1}{\nolinkurl{https://arxiv.org/abs/#1}}}

\bibitem[{{Abramowitz} \& {Stegun}(1965)}]{Abramovitz-Stegun:1965}
{Abramowitz}, M., \& {Stegun}, I.~A. 1965, Handbook of Mathematical Functions:
  with Formulas, Graphs, and Mathematical Tables. (Dover Publications, New
  York; revised edition)

\bibitem[{{Beichman} {et~al.}(2020){Beichman}, {Ygouf}, {Llop Sayson}, {Mawet},
  {Yung}, {Choquet}, {Kervella}, {Boccaletti}, {Belikov}, {Lissauer},
  {Quarles}, {Lagage}, {Dicken}, {Hu}, {Mennesson}, {Ressler}, {Serabyn},
  {Krist}, {Bendek}, {Leisenring}, \& {Pueyo}}]{Beichman-etal:2020}
{Beichman}, C., {Ygouf}, M., {Llop Sayson}, J., {et~al.} 2020, \pasp, 132,
  015002

\bibitem[{Bellm {et~al.}(2018)Bellm, Kulkarni, Graham, Dekany, Smith, Riddle,
  Masci, Helou, Prince, Adams, Barbarino, Barlow, Bauer, Beck, Belicki, Biswas,
  Blagorodnova, Bodewits, Bolin, Brinnel, Brooke, Bue, Bulla, Burruss, Cenko,
  Chang, Connolly, Coughlin, Cromer, Cunningham, De, Delacroix, Desai, Duev,
  Eadie, Farnham, Feeney, Feindt, Flynn, Franckowiak, Frederick, Fremling,
  Gal-Yam, Gezari, Giomi, Goldstein, Golkhou, Goobar, Groom, Hacopians, Hale,
  Henning, Ho, Hover, Howell, Hung, Huppenkothen, Imel, Ip, Ivezi{\'{c}},
  Jackson, Jones, Juric, Kasliwal, Kaspi, Kaye, Kelley, Kowalski, Kramer,
  Kupfer, Landry, Laher, Lee, Lin, Lin, Lunnan, Giomi, Mahabal, Mao, Miller,
  Monkewitz, Murphy, Ngeow, Nordin, Nugent, Ofek, Patterson, Penprase, Porter,
  Rauch, Rebbapragada, Reiley, Rigault, Rodriguez, van Roestel, Rusholme, van
  Santen, Schulze, Shupe, Singer, Soumagnac, Stein, Surace, Sollerman, Szkody,
  Taddia, Terek, Sistine, van Velzen, Vestrand, Walters, Ward, Ye, Yu, Yan, \&
  Zolkower}]{Bellm-etal:2018}
Bellm, E.~C., Kulkarni, S.~R., Graham, M.~J., {et~al.} 2018, PASP, 131, 018002

\bibitem[{{Born} \& {Wolf}(1999)}]{Born-Wolf:1999}
{Born}, M., \& {Wolf}, E. 1999, Principles of Optics: Electromagnetic Theory of
  Propagation, Interference and Diffraction of Light (Cambridge, England: 7-th
  edition. Cambridge University Press)

\bibitem[{Bovy {et~al.}(2009)Bovy, Hogg, \& Roweis}]{Bovy:2009}
Bovy, J., Hogg, D.~W., \& Roweis, S.~T. 2009, Astrophys. J., 700, 1794

\bibitem[{{Calchi Novati} \& {Scarpetta}(2016)}]{Calchi-Novati:2016}
{Calchi Novati}, S., \& {Scarpetta}, G. 2016, \apj, 824, 109

\bibitem[{{Carri{\'o}n-Gonz{\'a}lez} {et~al.}(2021){Carri{\'o}n-Gonz{\'a}lez},
  {Garc{\'\i}a Mu{\~n}oz}, {Santos}, {Cabrera}, {Csizmadia}, \&
  {Rauer}}]{Carrion-Gonzalez-eat:2021}
{Carri{\'o}n-Gonz{\'a}lez}, {\'O}., {Garc{\'\i}a Mu{\~n}oz}, A., {Santos},
  N.~C., {et~al.} 2021, \aap, 651, A7

\bibitem[{Carri{\'{o}}n-Gonz{\'{a}}lez
  {et~al.}(2021)Carri{\'{o}}n-Gonz{\'{a}}lez, Mu{\~{n}}oz, Santos, Cabrera,
  Csizmadia, \& Rauer}]{CarrionGonzalez-etal:2021}
Carri{\'{o}}n-Gonz{\'{a}}lez, {\'{O}}., Mu{\~{n}}oz, A.~G., Santos, N.~C.,
  {et~al.} 2021, A\&A, 651, A7

\bibitem[{Cassan \& Ranc(2016)}]{Cassan-Ranc:2016}
Cassan, A., \& Ranc, C. 2016, MNRAS, 458, 2074

\bibitem[{{Cassan} {et~al.}(2022){Cassan}, {Ranc}, {Absil}, {Wyrzykowski},
  {Rybicki}, {Bachelet}, {Le Bouquin}, {Hundertmark}, {Street}, {Surdej},
  {Tsapras}, {Wambsganss}, \& {Wertz}}]{Cassan-etal:2022}
{Cassan}, A., {Ranc}, C., {Absil}, O., {et~al.} 2022, Nature Astronomy, 6, 121

\bibitem[{Delplancke {et~al.}(2001)Delplancke, G{\'{o} }rski, \&
  Richichi}]{Delplancke:2001}
Delplancke, F., G{\'{o} }rski, K.~M., \& Richichi, A. 2001, Astron. Astrophys.,
  375, 701

\bibitem[{{Di Stefano}(2008)}]{DiStefano:2008}
{Di Stefano}, R. 2008, \apj, 684, 59

\bibitem[{Einstein(1936)}]{Einstein:1936}
Einstein, A. 1936, Science, 84, 506

\bibitem[{{Faria} {et~al.}(2022){Faria}, {Su\'arez Mascare\~no}, {Figueira},
  {Silva}, {Damasso}, {Demangeon}, {Pepe}, {Santos}, {Rebolo}, {Cristiani},
  {Adibekyan}, {Alibert}, {Allart}, {Barros}, {Cabral}, {D\'{}Odorico}, {Di
  Marcantonio}, {Dumusque}, {Ehrenreich}, {Gonz\'alez Hern\'andez}, {Hara},
  {Lillo-Box}, {Lo Curto}, {Lovis}, {Martins}, {M\'egevand}, {Mehner},
  {Micela}, {Molaro}, {Nunes}, {Pall\'e}, {Poretti}, {Sousa}, {Sozzetti},
  {Tabernero}, {Udry}, \& {Zapatero Osorio}}]{Faria-etal:2022}
{Faria}, J.~P., {Su\'arez Mascare\~no}, A., {Figueira}, P., {et~al.} 2022,
  \aap, 658, A115

\bibitem[{{Gaudi}(2012)}]{Gaudi:2012}
{Gaudi}, B.~S. 2012, Annual Rev. Astron. Astrophys., 50, 411

\bibitem[{{Gould}(1996)}]{Gould:1996}
{Gould}, A. 1996, PASP, 108, 465

\bibitem[{{Gould}(1997)}]{Gould:1997}
---. 1997, ApJ, 480, 188

\bibitem[{{Grudi{\'c}} {et~al.}(2019){Grudi{\'c}}, {Hopkins}, {Quataert}, \&
  {Murray}}]{Grudic-etal:2019}
{Grudi{\'c}}, M.~Y., {Hopkins}, P.~F., {Quataert}, E., \& {Murray}, N. 2019,
  \mnras, 483, 5548

\bibitem[{{Heap} {et~al.}(2017){Heap}, {Szalay}, \& {WFIRST Science Archive
  Team}}]{2017IAUS..325..373H}
{Heap}, S.~R., {Szalay}, A.~S., \& {WFIRST Science Archive Team}. 2017, in
  Astroinformatics, ed. M.~{Brescia}, S.~G. {Djorgovski}, E.~D. {Feigelson},
  G.~{Longo}, \& S.~{Cavuoti}, Vol. 325, 373--378

\bibitem[{{Hopkins} {et~al.}(2010){Hopkins}, {Murray}, {Quataert}, \&
  {Thompson}}]{Hopkins-2010}
{Hopkins}, P.~F., {Murray}, N., {Quataert}, E., \& {Thompson}, T.~A. 2010,
  MNRAS, 401, L19

\bibitem[{{Ji} {et~al.}(2022){Ji}, {Li}, {Zhang}, {Fang}, {Li}, {Wang}, {Cao},
  {Deng}, {Li}, {Xian}, {Gao}, {Zhang}, {Li}, {Liu}, {Qi}, {Jin}, {Liu},
  {Chen}, {Li}, {Dong}, \& {Zhu}}]{Ji-etal:2022}
{Ji}, J.-H., {Li}, H.-T., {Zhang}, J.-B., {et~al.} 2022, Res. in Astron.
  Astrophys., 22, 072003

\bibitem[{Kamphues(2016)}]{TMT2016}
Kamphues, F. 2016, Mikroniek: Magazine of the Dutch Society for Precision
  Engineering, 6, 10

\bibitem[{{Kasper} {et~al.}(2010){Kasper}, {Beuzit}, {Verinaud}, {Gratton},
  {Kerber}, {Yaitskova}, {Boccaletti}, {Thatte}, {Schmid}, {Keller}, {Baudoz},
  {Abe}, {Aller-Carpentier}, {Antichi}, {Bonavita}, {Dohlen}, {Fedrigo},
  {Hanenburg}, {Hubin}, {Jager}, {Korkiakoski}, {Martinez}, {Mesa}, {Preis},
  {Rabou}, {Roelfsema}, {Salter}, {Tecza}, \& {Venema}}]{2010SPIE.7735E..2EK}
{Kasper}, M., {Beuzit}, J.-L., {Verinaud}, C., {et~al.} 2010, in Society of
  Photo-Optical Instrumentation Engineers (SPIE) Conference Series, Vol. 7735,
  Ground-based and Airborne Instrumentation for Astronomy III, ed. I.~S.
  {McLean}, S.~K. {Ramsay}, \& H.~{Takami}, 77352E

\bibitem[{{Kervella} {et~al.}(2016){Kervella}, {Mignard}, {M{\'e}rand}, \&
  {Th{\'e}venin}}]{Kervella2016}
{Kervella}, P., {Mignard}, F., {M{\'e}rand}, A., \& {Th{\'e}venin}, F. 2016,
  \aap, 594, A107

\bibitem[{Kim {et~al.}(2010)Kim, Park, Lee, Yuk, Han, O'Brien, Gould, Lee, \&
  Kim}]{Kim-etal:2010}
Kim, S.-L., Park, B.-G., Lee, C.-U., {et~al.} 2010, in Ground-based and
  Airborne Telescopes III, ed. L.~M. Stepp, R.~Gilmozzi, \& H.~J. Hall, Vol.
  7733, International Society for Optics and Photonics (SPIE), 77333F

\bibitem[{{K{\i}ro{\u{g}}lu} {et~al.}(2022){K{\i}ro{\u{g}}lu}, {Weatherford},
  {Kremer}, {Ye}, {Fragione}, \& {Rasio}}]{Kiroglu-etal:2022}
{K{\i}ro{\u{g}}lu}, F., {Weatherford}, N.~C., {Kremer}, K., {et~al.} 2022,
  \apj, 928, 181

\bibitem[{{Kl{\"u}ter} {et~al.}(2018){Kl{\"u}ter}, {Bastian}, {Demleitner}, \&
  {Wambsganss}}]{Kluter-etal:2018}
{Kl{\"u}ter}, J., {Bastian}, U., {Demleitner}, M., \& {Wambsganss}, J. 2018,
  \aap, 620, A175

\bibitem[{Liebes(1964)}]{Liebes:1964}
Liebes, S. 1964, Phys. Rev., 133, B835

\bibitem[{{Lindegren} {et~al.}(2021){Lindegren}, {Bastian}, {Biermann},
  {Bombrun}, {de Torres}, {Gerlach}, {Geyer}, {Hern\'andez}, {Hilger}, {Hobbs},
  {Klioner}, {Lammers}, {McMillan}, {Ramos-Lerate}, {Steidelm\"uller},
  {Stephenson}, \& {van Leeuwen}}]{Lindegren-eta:2021}
{Lindegren}, L., {Bastian}, U., {Biermann}, M., {et~al.} 2021, \aap, 649, A4

\bibitem[{{Mao}(2012)}]{Mao:2012}
{Mao}, S. 2012, Res. Astron. Astrophys., 12, 947

\bibitem[{{Mustill} {et~al.}(2018){Mustill}, {Davies}, \&
  {Lindegren}}]{Mustill-etal:2018}
{Mustill}, A.~J., {Davies}, M.~B., \& {Lindegren}, L. 2018, \aap, 617, A135

\bibitem[{{Ochsenbein} \& et~al.(2019)}]{vizier}
{Ochsenbein}, F., \& et~al. 2019, {The VizieR database of astronomical
  catalogues}, \dodoi{10.26093/cds/vizier}

\bibitem[{{Paczy{\'{n}}ski}(1986)}]{Paczynski:1986}
{Paczy{\'{n}}ski}, B. 1986, ApJ, 301, 503

\bibitem[{{Paczy{\'{n}}ski}(1996)}]{Paczynski:1996a}
---. 1996, Ann. Rev. Astron. Astrophys., 34, 419

\bibitem[{{Rodriguez} {et~al.}(2022){Rodriguez}, {Mr{\'o}z}, {Kulkarni},
  {Andreoni}, {Bellm}, {Dekany}, {Drake}, {Duev}, {Graham}, {Masci}, {Prince},
  {Riddle}, \& {Shupe}}]{Rodriguez-eatal:2022}
{Rodriguez}, A.~C., {Mr{\'o}z}, P., {Kulkarni}, S.~R., {et~al.} 2022, \apj,
  927, 150

\bibitem[{{Schneider} {et~al.}(1992){Schneider}, {Ehlers}, \&
  {Falco}}]{Schneider-Ehlers-Falco:1992}
{Schneider}, P.~S., {Ehlers}, J., \& {Falco}, E. 1992, Gravitational Lenses
  (Springer-Verlag Berlin Heidelberg)

\bibitem[{{Shao} {et~al.}(2022){Shao}, {Zhai}, {Nemati}, {Hahn}, {Trahan}, \&
  {Turyshev}}]{Shao-etal:2022}
{Shao}, M., {Zhai}, C., {Nemati}, B., {et~al.} 2022, Micro-arcsecond Astrometry
  Technology: Detector and Field Distortion Calibration, arXiv:2212.08129
  [astro-ph.IM]

\bibitem[{Shao {et~al.}(2018)Shao, Turyshev, Bendek, Fischer, Guyon, McArthur,
  Muterspaugh, Zhai, \& Boehm}]{Shao-etal:2018}
Shao, M., Turyshev, S.~G., Bendek, E., {et~al.} 2018, Precision Space
  Astrometry as a Tool to Find Earth-like Exoplanets, arXiv:1803.03732
  [astro-ph.IM]

\bibitem[{{Smart} {et~al.}(2021){Smart}, {Sarro}, {Rybizki}, Reyle, Robin,
  Hambly, Abbas, Barstow, de~Bruijne, \& Bucciarelli}]{Smart-etal:2021}
{Smart}, R.~L., {Sarro}, L.~M., {Rybizki}, J., {et~al.} 2021, A\&A, 649, A6

\bibitem[{{Stobie} {et~al.}(1995){Stobie}, {Glass}, \& {Buckley}}]{SALT1995}
{Stobie}, R.~S., {Glass}, I.~S., \& {Buckley}, D. A.~H. 1995, {South African
  Journal of Science}, 91, 30

\bibitem[{{Street} \& {Bachelet}(2019)}]{Street-Bachelet:2019}
{Street}, R.~A., \& {Bachelet}, E. 2019, AJ, 158, 110

\bibitem[{{Thatte} {et~al.}(2021){Thatte}, {Tecza}, {Schnetler}, {Neichel},
  {Melotte}, {Fusco}, {Ferraro-Wood}, {Clarke}, {Bryson}, {O'Brien}, {Mateo},
  {Garcia Lorenzo}, {Evans}, {Bouch{\'e}}, {Arribas}, \& {HARMONI
  Consortium}}]{2021Msngr.182....7T}
{Thatte}, N., {Tecza}, M., {Schnetler}, H., {et~al.} 2021, The Messenger, 182,
  7

\bibitem[{{Toth} \& {Turyshev}(2021)}]{Toth-Turyshev:2020}
{Toth}, V.~T., \& {Turyshev}, S.~G. 2021, Phys. Rev. D, 103, 124038

\bibitem[{{Toth} \& {Turyshev}(2022)}]{SGL2021h}
---. 2022, Int. J. Geom. Methods Mod. Phys., 19, 2250201

\bibitem[{{Traub} \& {Oppenheimer}(2010)}]{Traub-Oppenheimer:2010}
{Traub}, W.~A., \& {Oppenheimer}, B.~R. 2010, in Exoplanets, ed. S.~{Seager}
  (``Exoplanets'', University of Arizona Press, Tucson, AZ), 111--156

\bibitem[{Tsapras {et~al.}(2009)Tsapras, Street, Horne, Snodgrass, Dominik,
  Allan, Steele, Bramich, Saunders, Rattenbury, Mottram, Fraser, Clay,
  Burgdorf, Bode, Lister, Hawkins, Beaulieu, Fouqu{\'{e} }, Albrow, Menzies,
  Cassan, \& Dominis-Prester}]{Tsapras-etal:2009}
Tsapras, Y., Street, R., Horne, K., {et~al.} 2009, Astronomische Nachrichten,
  330, 4

\bibitem[{{Turyshev} \& {Toth}(2017)}]{Turyshev-Toth:2017}
{Turyshev}, S.~G., \& {Toth}, V.~T. 2017, Phys. Rev. D, 96, 024008

\bibitem[{{Turyshev} \& {Toth}(2019)}]{Turyshev-Toth:2019-extend}
---. 2019, Phys. Rev. D, 100, 084018

\bibitem[{{Turyshev} \& {Toth}(2020{\natexlab{a}})}]{Turyshev-Toth:2020-extend}
---. 2020{\natexlab{a}}, Phys. Rev. D, 102, 024038

\bibitem[{{Turyshev} \& {Toth}(2020{\natexlab{b}})}]{Turyshev-Toth:2019-blur}
---. 2020{\natexlab{b}}, Phys. Rev. D, 101, 044025

\bibitem[{{Turyshev} \& {Toth}(2020{\natexlab{c}})}]{Turyshev-Toth:2019-image}
---. 2020{\natexlab{c}}, Phys. Rev. D, 101, 044048

\bibitem[{{Turyshev} \&
  {Toth}(2021{\natexlab{a}})}]{Turyshev-Toth:2021-multipoles}
---. 2021{\natexlab{a}}, Phys. Rev. D, 103, 064076

\bibitem[{{Turyshev} \&
  {Toth}(2021{\natexlab{b}})}]{Turyshev-Toth:2021-imaging}
---. 2021{\natexlab{b}}, Phys. Rev. D, 104, 044032

\bibitem[{{Turyshev} \&
  {Toth}(2021{\natexlab{c}})}]{Turyshev-Toth:2021-quartic}
---. 2021{\natexlab{c}}, Phys. Rev. D, 104, 124033

\bibitem[{{Turyshev} \&
  {Toth}(2021{\natexlab{d}})}]{Turyshev-Toth:2021-caustics}
---. 2021{\natexlab{d}}, Phys. Rev. D, 104, 024019

\bibitem[{{Turyshev} \&
  {Toth}(2021{\natexlab{e}})}]{Turyshev-Toth:2021-all-regions}
---. 2021{\natexlab{e}}, Phys. Rev. D, 104, 044013

\bibitem[{{Turyshev} \&
  {Toth}(2022{\natexlab{a}})}]{Turyshev-Toth:2021-STF-moments}
---. 2022{\natexlab{a}}, Phys. Rev. D, 105, 024022

\bibitem[{{Turyshev} \&
  {Toth}(2022{\natexlab{b}})}]{Turyshev-Toth:2022-broad_SNR}
---. 2022{\natexlab{b}}, MNRAS, 515, 6122

\bibitem[{{Turyshev} \&
  {Toth}(2022{\natexlab{c}})}]{Turyshev-Toth:2022-mono_SNR}
---. 2022{\natexlab{c}}, Phys. Rev. D, 106, 044059

\bibitem[{{Turyshev} \&
  {Toth}(2022{\natexlab{d}})}]{Turyshev-Toth:2022-wobbles}
---. 2022{\natexlab{d}}, Phys. Rev. D, 105, 044012

\bibitem[{{Turyshev} \& {Toth}(2022{\natexlab{e}})}]{SGL2021g}
---. 2022{\natexlab{e}}, MNRAS, 513, 5355

\bibitem[{{Udalski} {et~al.}(1992){Udalski}, {Szymanski}, {Kaluzny}, {Kubiak},
  \& {Mateo}}]{Udalski:1992}
{Udalski}, A., {Szymanski}, M., {Kaluzny}, J., {Kubiak}, M., \& {Mateo}, M.
  1992, Acta Astronomica, 42, 253

\bibitem[{Wizinowich {et~al.}(2000)Wizinowich, Acton, Shelton, Stomski,
  Gathright, Ho, Lupton, Tsubota, Lai, Max, Brase, An, Avicola, Olivier, Gavel,
  Macintosh, Ghez, \& Larkin}]{Wizinowich_2000}
Wizinowich, P., Acton, D.~S., Shelton, C., {et~al.} 2000, PASP, 112, 315

\bibitem[{{Zhu} \& {Gould}(2016)}]{Zhu-Gould:2016}
{Zhu}, W., \& {Gould}, A. 2016, J. Korean Astron. Soc., 49, 93

\bibitem[{{Zwicky}(1937)}]{Zwicky:1937}
{Zwicky}, F. 1937, Phys. Rev., 51, 290

\end{thebibliography}
\bibliographystyle{aasjournal}

\end{document}